\documentclass[aps,prb,twocolumn,superscriptaddress,groupedaddress,a4paper]{revtex4}  % Uncomment for review/submission version
\pdfoutput=1
\usepackage{graphicx}
\usepackage{dcolumn}
\usepackage{bm}
\usepackage{amssymb}
\usepackage{amsmath}
\usepackage{subfigure}
\usepackage{color}
\newcommand{\kdotp}{$\textbf{k}\cdot\textbf{p}$}
\newcommand{\spss}{$sp^{3}s^{*}$}
\newcommand{\GaNAs}{GaN$_{y}$As$_{1-y}$}
\newcommand{\GaNP}{GaN$_{y}$P$_{1-y}$}
\newcommand{\InGaNAs}{(In)GaN$_{y}$As$_{1-y}$}
\newcommand{\GaBiAs}{GaBi$_{x}$As$_{1-x}$}
\newcommand{\GaBiP}{GaBi$_{x}$P$_{1-x}$}
\newcommand{\GaBiNAs}{GaBi$_{x}$N$_{y}$As$_{1-x-y}$}
\newcommand{\Eg}{$E_{g}$}
\newcommand{\DeltaSO}{$\Delta_{\scalebox{0.6}{\textrm{SO}}}$}
\newcommand{\EBi}{$E_{\scalebox{0.6}{\textrm{Bi}}}$}
\newcommand{\betaBi}{$\beta_{\scalebox{0.6}{\textrm{Bi}}}$}
\newcommand{\psiBi}{$\vert \psi_{\scalebox{0.6}{\textrm{Bi}}} \rangle$}
\newcommand{\EN}{$E_{\scalebox{0.6}{\textrm{N}}}$}

\begin{document}

\widetext
\title{Derivation of 12- and 14-band \kdotp\, Hamiltonians for\\dilute bismide and bismide-nitride semiconductors}

%%%%%%%%%%%%%%%%%%%%%%%%%%%%%%%%%%%%%%%%
%%%% Authors, affiliations and date %%%%
%%%%%%%%%%%%%%%%%%%%%%%%%%%%%%%%%%%%%%%%

\author{Christopher A. Broderick}
\email{chris.broderick@tyndall.ie} % Email address must be kept before affiliation(s)
\affiliation{Tyndall National Institute, Lee Maltings, Dyke Parade, Cork, Ireland}
\affiliation{Department of Physics, University College Cork, Cork, Ireland}

\author{Muhammad Usman}
\affiliation{Tyndall National Institute, Lee Maltings, Dyke Parade, Cork, Ireland}

\author{Eoin P. O'Reilly}
\affiliation{Tyndall National Institute, Lee Maltings, Dyke Parade, Cork, Ireland}
\affiliation{Department of Physics, University College Cork, Cork, Ireland}

\vskip 0.25cm

\date{\today}

%%%%%%%%%%%%%%%%%%%%%%%%%%%%%%%%%%%%%%%%%%%%%
%%%% Abstract, keywords and PACS numbers %%%%
%%%%%%%%%%%%%%%%%%%%%%%%%%%%%%%%%%%%%%%%%%%%%

\begin{abstract}

Using an \spss\, tight-binding model we demonstrate how the observed strong bowing of the band gap and spin-orbit-splitting with increasing Bi composition in the dilute bismide alloy \GaBiAs\, can be described in terms of a band-anticrossing interaction between the extended states of the GaAs valence band edge and highly localised Bi-related resonant states lying below the GaAs valence band edge. We derive a 12-band \kdotp\, Hamiltonian to describe the band structure of \GaBiAs\, and show that this model is in excellent agreement with full tight-binding calculations of the band structure in the vicinity of the band edges, as well as with experimental measurements of the band gap and spin-orbit-splitting across a large composition range. Based on a tight-binding model of \GaBiNAs\, we show that to a good approximation N and Bi act independently of one another in disordered \GaBiNAs\, alloys, indicating that a simple description of the band structure is possible. We present a 14-band \kdotp\, Hamiltonian for ordered \GaBiNAs\, crystals which reproduces accurately the essential features of full tight-binding calculations of the band structure in the vicinity of the band edges. The \kdotp\, models we present here are therefore ideally suited to the simulation of the optoelectronic properties of these novel III-V semiconductor alloys.

\end{abstract}

% \keywords{XXXX}
% \pacs{XXXX}

\maketitle

%%%%%%%%%%%%%%%%%%%%%%
%%%% Introduction %%%%
%%%%%%%%%%%%%%%%%%%%%%

\section{Introduction}
\label{sec:introduction}

Highly mismatched semiconductor alloys such as \GaNAs\, and \GaBiAs\, have attracted considerable interest in recent years, both from a fundamental perspective and also because of their potential device applications \cite{Reilly_SST_2009,Henini}. When a small fraction of As is replaced by nitrogen (N) in GaAs, the band gap (\Eg) initially decreases rapidly, by $\sim 150$ meV when 1\% of As is replaced by N \cite{Shan_PRL_1999}.

Similar behavior has been experimentally observed in \GaBiAs, where \Eg\, decreases initially by $\sim 90$ meV per \% of bismuth (Bi) replacing As \cite{Alberi_PRB_2007,Tixier_APL_2005}. Additionally, recent experiments \cite{Batool_JAP_2012,Fluegel_PRL_2006} have also revealed the presence of a large bowing of the spin-orbit-splitting (SO) energy (\DeltaSO), which increases strongly with increasing Bi composition.

It was shown recently using photo-modulated reflectance (PR) spectroscopy that increasing the Bi composition in \GaBiAs\, leads to the onset of an \Eg $<$ \DeltaSO\, regime in the alloy \cite{Batool_JAP_2012,Usman_PRB_2013}. This regime is of interest for the design of highly efficient and thermally stable optoelectronic devices since it opens up the possibility of suppression of the dominant non-radiative Auger recombination pathway, the presence of which dominates the threshold current and degrades temperature stability of III-V lasers operating at telecommunication wavelengths \cite{Phillips_IEEEJSTQE_1999,Sweeney_ICTON_2011}. Investigation of dilute bismide and bismide-nitride alloys of GaAs is therefore highly promising for potential device applications.

The extreme band gap bowing observed in \InGaNAs\, has been well explained in terms of a band-anticrossing (BAC) interaction between two levels, one at energy $E_{\scalebox{0.6}{\textrm{CB}}}$ associated with the extended conduction band edge (CBE) states of the host (In)GaAs matrix, and the second at energy \EN\, associated with the highly localised N-related resonant impurity states in the alloy. In this simple model, the CBE energy of the N-containing alloy is given by the lower eigenvalue, $E^{\scalebox{0.6}{\textrm{CB}}}_{-}$, of the 2-band Hamiltonian \cite{Shan_PRL_1999}:

% Equation 1

\begin{equation}
	\left(
	\begin{array}{cc}
		E_{\scalebox{0.6}{\textrm{N}}}  & V_{\scalebox{0.6}{\textrm{Nc}}} \\
		V_{\scalebox{0.6}{\textrm{Nc}}} & E_{\scalebox{0.6}{\textrm{CB}}}
	\end{array}
	\right)
	\label{eq:2_band_BAC}
\end{equation}

\noindent
where $V_{\scalebox{0.6}{\textrm{Nc}}}$ is the N composition dependent matrix element describing the interaction between $E_{\scalebox{0.6}{\textrm{CB}}}$ and \EN, usually taken to vary with N composition, $y$, as $V_{\scalebox{0.6}{\textrm{Nc}}} = \beta_{\scalebox{0.6}{\textrm{N}}} \sqrt{y}$.

Bismuth, being the heaviest stable group V element, is significantly larger and more electropositive than As. It should therefore be expected that any Bi-related impurity levels should either lie below or close in energy to the valence band edge (VBE) and that, if an anticrossing interaction occurs, it will occur between the Bi-related impurity levels and the VBE of the host GaAs matrix.

Recently the presence of such an interaction has been proposed \cite{Alberi_PRB_2007,Alberi_APL_2007,Usman_PRB_2011}, but there has been controversy as to whether or not a BAC-like model can be applied in the case of the dilute bismides. This controversy is due in part to the absence in PR spectra of Bi-related features in the \GaBiAs\, valence band which, from previous studies of \GaNAs, one would expect to be present in the case of a BAC interaction. In addition, analysis of the band structure of small Ga$_{M}$Bi$_{1}$As$_{M-1}$ supercells using density functional theory (DFT) calculations failed to find evidence of Bi-related resonant states immediately below the valence band maximum \cite{Deng_PRB_2010}.

In Ref.~\onlinecite{Usman_PRB_2011} we presented a nearest-neighbor \spss\, tight-binding (TB) model to describe the electronic structure of dilute bismide alloys of GaAs and GaP. The \spss\, model agrees well both with experimental measurements and pseudopotential calculations\cite{Zhang_PRB_2005} of the variation of \Eg\, and \DeltaSO\, with Bi composition in \GaBiAs\, over the investigated composition range.

Calculations based on our TB model reproduce the experimentally observed transition to an \Eg \, $<$ \DeltaSO\, regime in both free-standing and strained alloys for $x \approx 10$\% \cite{Usman_PRB_2011} and $x \approx 9$\% \cite{Batool_JAP_2012,Usman_PRB_2013} respectively. These calculations also agreed with the previously published conclusions of the pseudopotential calculations of Ref.~\onlinecite{Zhang_PRB_2005}, namely that Bi forms a resonant state below the GaAs VBE and that the alloy VBE is derived predominantly from the host VBE as opposed to a Bi bound state.

Detailed analysis of our TB calculations support the presence of a BAC interaction between the GaAs VBE and lower-lying Bi-related impurity states which are resonant with the GaAs valence band. We show, in agreement with our previous work, that the energy of the Bi resonant state in Ga$_M$Bi$_1$As$_{M-1}$ ordered supercell calculations shifts down significantly in energy for small values of $M$, thereby accounting for the failure to observe such a state in previous small supercell calculations\cite{Deng_PRB_2010}. In addition, our analysis indicates that the failure to observe any Bi-related features below the \GaBiAs\, VBE in PR spectra results from the broadening of the Bi impurity states by the large density of host valence states with which they are resonant \cite{Usman_PRB_2011}.

The realisation recently of the first electrically pumped dilute bismide quantum well laser \cite{Ludewig_APL_2013} indicates the progress which has been made in the material growth and understanding of this highly-mismatched material system. From a theoretical standpoint, this milestone mandates the development of models suited to the description of the optoelectronic characteristics of dilute bismide-based quantum well lasers. While our atomistic TB model has revealed in detail the effects of Bi on the electronic structure of GaAs, this comes at significant computational cost. Atomistic theoretical models have been applied with success to analyse quantum dot heterostructures, but due to the additional degrees of freedom present in a quantum well (where the Brillouin zone is two-dimensional, as opposed to the zero-dimensional quantum dot case) continuum models of the band structure are favoured for the description of the optoelectronic properties. The reason for this is that the additional degrees of freedom, specifically, the carrier wave vectors, must be integrated out in order to calculate key physical quantities such as the material gain, which is computationally expensive even in the continuum case. The validity of the continuum approach to calculating quantum well band structure has been thoroughly verified with \kdotp-based many-body calculations \cite{Buckers_PSSB_2010} having been employed to accurately describe the gain characteristics of, amongst others, GaInP \cite{Chow_APL_1997}, InGaN \cite{Lermer_APL_2011} and dilute nitride GaInNAs \cite{Thranhardt_APL_2005} quantum well lasers. These factors strongly motivate the development of simple and accurate models of the GaBi$_{x}$As$_{1-x}$ band structure, which can then be applied to the study of the electronic and optical properties of dilute bismide materials and devices.

In this work, we apply the TB model to derive a \kdotp\, Hamiltonian suitable to describe the band dispersion of \GaBiAs. Based on ordered supercell calculations using the TB model, we derive a 12-band \kdotp\, model for \GaBiAs\, which includes BAC interactions between the GaAs VBE and Bi impurity states  which are resonant with the GaAs valence band. The model is shown to be in excellent agreement with full TB calculations on ordered and large, disordered supercells, as well as with experimental measurements of the band gap and spin-orbit-splitting across the full composition range considered ($x = $ 0 -- 12\%) and, combined with a detailed TB-based analysis, demonstrates emphatically the applicability of the BAC model to \GaBiAs.

The dilute bismide-nitride alloy \GaBiNAs\, has also been identified as being very promising for the design of highly efficient optoelectronic devices, since it opens up several avenues by which the band structure can be engineered \cite{Broderick_SST_2012,Sweeney_JAP_2013}. In particular, \GaBiNAs\, is predicted to have a giant band gap bowing, allowing for emission at 1.3 and 1.55 $\mu$m on a GaAs substrate, as well as retaining the large spin-orbit-splitting bowing that is characteristic of \GaBiAs, thereby offering the possibility of suppressing the non-radiative CHSH Auger recombination process at longer wavelengths \cite{Sweeney_JAP_2013}. Also, incorporation of Bi and N in GaAs introduces compressive and tensile strain respectively when grown on a GaAs substrate, so that a high degree of strain engineering should be possible in \GaBiNAs\, when grown on GaAs, providing further opportunities for the manipulation of the electronic properties \cite{Janotti_PRB_2002}.

Based on our TB model for \GaBiNAs\, we show that the effects of Bi and N on the electronic structure of GaAs are largely independent of one another, both in ordered and disordered crystals \cite{Broderick_ICTON_2011,Usman_GaBiNAs_2013}. We can therefore derive a 14-band \kdotp\, model for \GaBiNAs\, which accurately describes the near zone centre band dispersion of ordered \GaBiNAs\, crystals. Finally we conclude our discussion of the \GaBiNAs\, band structure by presenting calculations of \Eg\, and \DeltaSO\, as a function of Bi ($0 \leq x \leq 12$\%) and N ($0 \leq y \leq 7$\%) composition. Our calculations show that a large wavelength range is accessible under low strain on a GaAs substrate and highlight the wide parameter space where the CHSH Auger-suppressing band structure condition \Eg \, $<$ \, \DeltaSO\, is fulfilled.

The remainder of this paper is organised as follows: In Section~\ref{sec:BAC_evidence} we demonstrate the presence of a BAC interaction in the valence band of \GaBiAs\, by using the TB model to construct and explore the character of the resonant state associated with an isolated Bi impurity in a series of ordered Ga$_{M}$Bi$_{1}$As$_{M-1}$ supercells. In Section~\ref{sec:12band} we derive a 12-band \kdotp\, Hamiltonian for \GaBiAs\, from the TB calculations and compare the resulting model to the results of full TB calculations for ordered and disordered \GaBiAs\, supercells. We present in Section~\ref{sec:14_band_model} a 14-band \kdotp\, model for \GaBiNAs. We first compare the results of the model to full TB supercell calculations, and then discuss some of the possibilities for band structure engineering offered by this novel quaternary alloy. Finally, in Section~\ref{sec:conclusions} we summarise and conclude.

%%%%%%%%%%%%%%%%%%%%%%%%%%%%%%%%%%%%%%%%%%%%%%
%%%% Section 2: Evidence of BAC in GaBiAs %%%%
%%%%%%%%%%%%%%%%%%%%%%%%%%%%%%%%%%%%%%%%%%%%%%

\section{Tight-binding analysis of band-anticrossing in G\lowercase{a}B\lowercase{i}$_{x}$A\lowercase{s}$_{1-x}$}
\label{sec:BAC_evidence}

We showed in Ref.~\onlinecite{Usman_PRB_2011} that the spectrum of fractional $\Gamma$ character, $f_{\Gamma}$, for the states in a supercell is a useful tool for analyzing the evolution of the electronic structure of dilute bismide alloys, where $f_{\Gamma}$ for a given state refers to the fraction of the state which can be projected onto the host matrix material band edge ($\Gamma$-point) states. By calculating this spectrum, $G_{\Gamma} (E)$, for a range of ordered \GaBiP\, supercells, we showed that an isolated, substitutional Bi atom acts as an impurity in GaP, giving rise to a highly-localised four-fold degenerate impurity level lying approximately 0.1 eV above the GaP VBE, in good agreement with experiment \cite{Trumbore_APL_1964}.

$G_{\Gamma} (E)$ is calculated in general by projecting a specific choice of host matrix (in this case, GaAs) band edge states at $\Gamma$ onto the full spectrum of levels in the alloy supercell. Using the subscripts $l,0$ and $k,1$ to denote the unperturbed host and Bi-containing alloy supercell states respectively, we obtain $G_{\Gamma}(E)$ by projecting the unperturbed Ga$_{M}$As$_{M}$ states at $\Gamma$, $\vert \psi_{l,0} \rangle$, onto the spectrum of a Ga$_{M}$Bi$_{L}$As$_{M-L}$ alloy supercell, $\left\lbrace E_{k}, \vert \psi_{k,1} \rangle \right\rbrace$:

% Equation 2

\begin{equation}
	G_{\Gamma} \left( E \right) = \sum_{k} \sum_{l=1}^{g(E_{l})} \vert \langle \psi_{k,1} \vert \psi_{l,0} \rangle \vert ^{2} \; T \left( E - E_{k} \right)
\end{equation}

\noindent
where $g(E_{l})$ is the degeneracy of the host band having energy $E_{l}$ at $\Gamma$ (so that $g(E_{l}) = 2, 4$ and 2 for the conduction, highest valence and SO bands, respectively) and we choose $T \left( E - E_{k} \right)$ so that $G_{\Gamma} (E_{k})$ has a value of unity for a doubly degenerate host matrix $\Gamma$ state at energy $E_{k}$. 

By considering the distribution of $G_{\Gamma} (E)$ for the GaP VBE (i.e. light-hole (LH) + heavy-hole (HH) band edges), in a series of ordered, cubic Ga$_{M}$Bi$_{1}$P$_{M-1}$ supercells we showed that this impurity level interacts with the GaP VBE via an anticrossing interaction which we found to vary with Bi composition as $V_{\scalebox{0.6}{\textrm{Bi}}} = \beta_{\scalebox{0.6}{\textrm{Bi}}} \sqrt{x}$.

A similar BAC interaction can be demonstrated in \GaBiAs, as we describe here by considering the $G_{\Gamma} (E)$ spectra of the GaAs VBE and the impurity state associated with a substitutional Bi atom in a series of ordered, cubic Ga$_{M}$Bi$_{1}$As$_{M-1}$ supercells containing $2M = 8N^{3}$ atoms, for $2 \leq N \leq 8$.
We proceed by inserting a single substitutional Bi impurity in a series of cubic $2M$-atom Ga$_{M}$As$_{M}$ supercells and calculate the $G_{\Gamma} (E)$ spectrum in each case by projecting the wave functions of the four-fold degenerate Ga$_{M}$As$_{M}$ VBE onto the full spectrum of the Bi-containing alloy \cite{Usman_PRB_2011}.

We can see from Fig.~\ref{fig:FGC_VBM_ordered_GaBiAs} that the GaAs VBE character in each supercell resides predominantly ($\gtrsim 90$\%) on the alloy VB maximum, with the remainder distributed over a series of lower-lying valence levels. This is in agreement with the pseudopotential calculations of Zhang, \emph{et al.} \cite{Zhang_PRB_2005}, namely that the \GaBiAs\, VBE is primarily derived from that of GaAs, with Bi forming a resonant impurity state lying below the GaAs VBE in energy.

In the BAC model, the \GaBiAs\, VBE is an admixture of the GaAs VBE and the impurity state associated with a Bi impurity. Writing the alloy VBE as a linear combination in this manner we obtain an expression for the four Bi-related states which mix with the GaAs VBE:

% Figure 1

\begin{figure*}[tb!]
	\hspace{-0.75cm} \subfigure{ \includegraphics[scale=0.7]{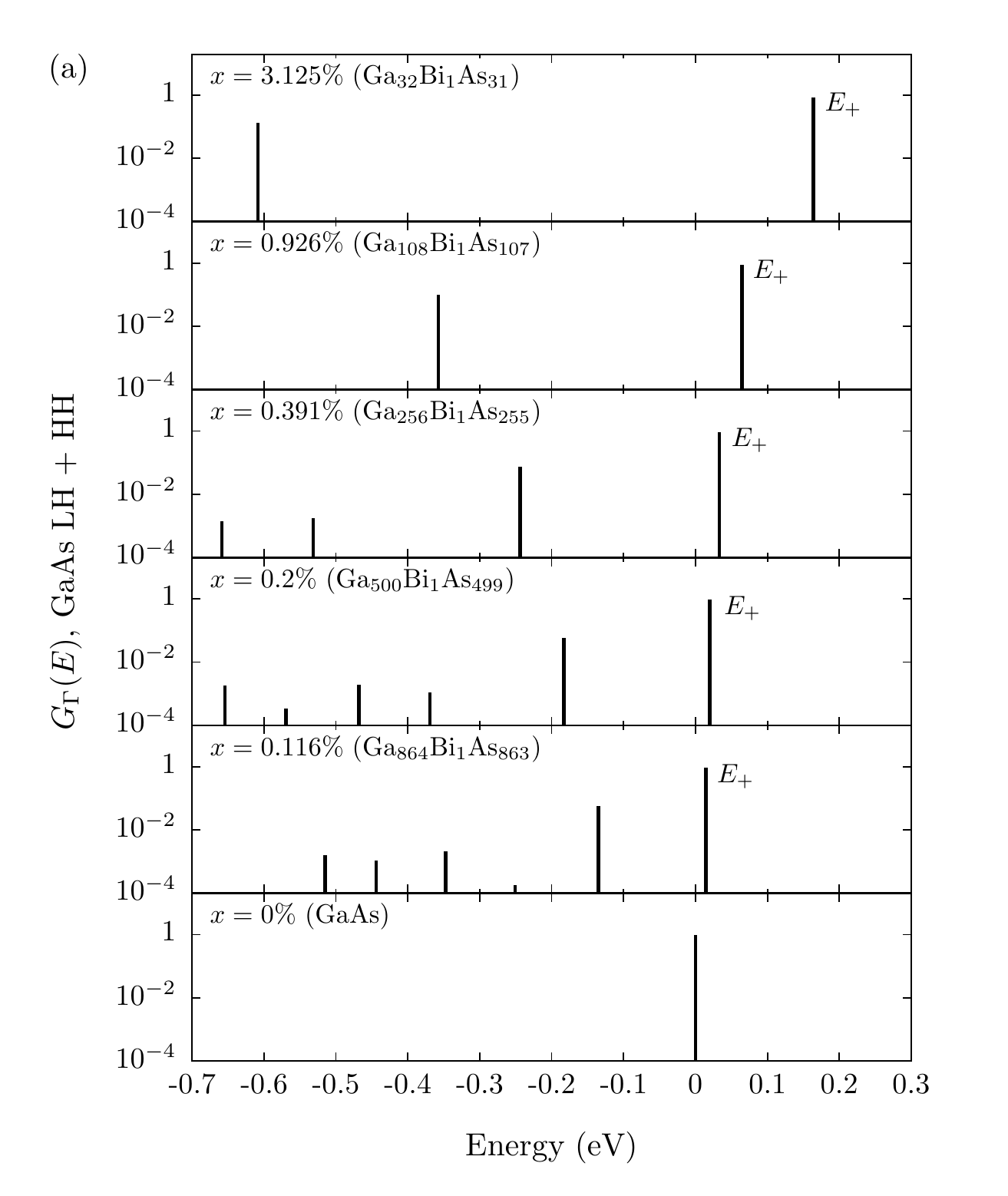} \label{fig:FGC_VBM_ordered_GaBiAs} }
	\hspace{-1.00cm} \subfigure{ \includegraphics[scale=0.7]{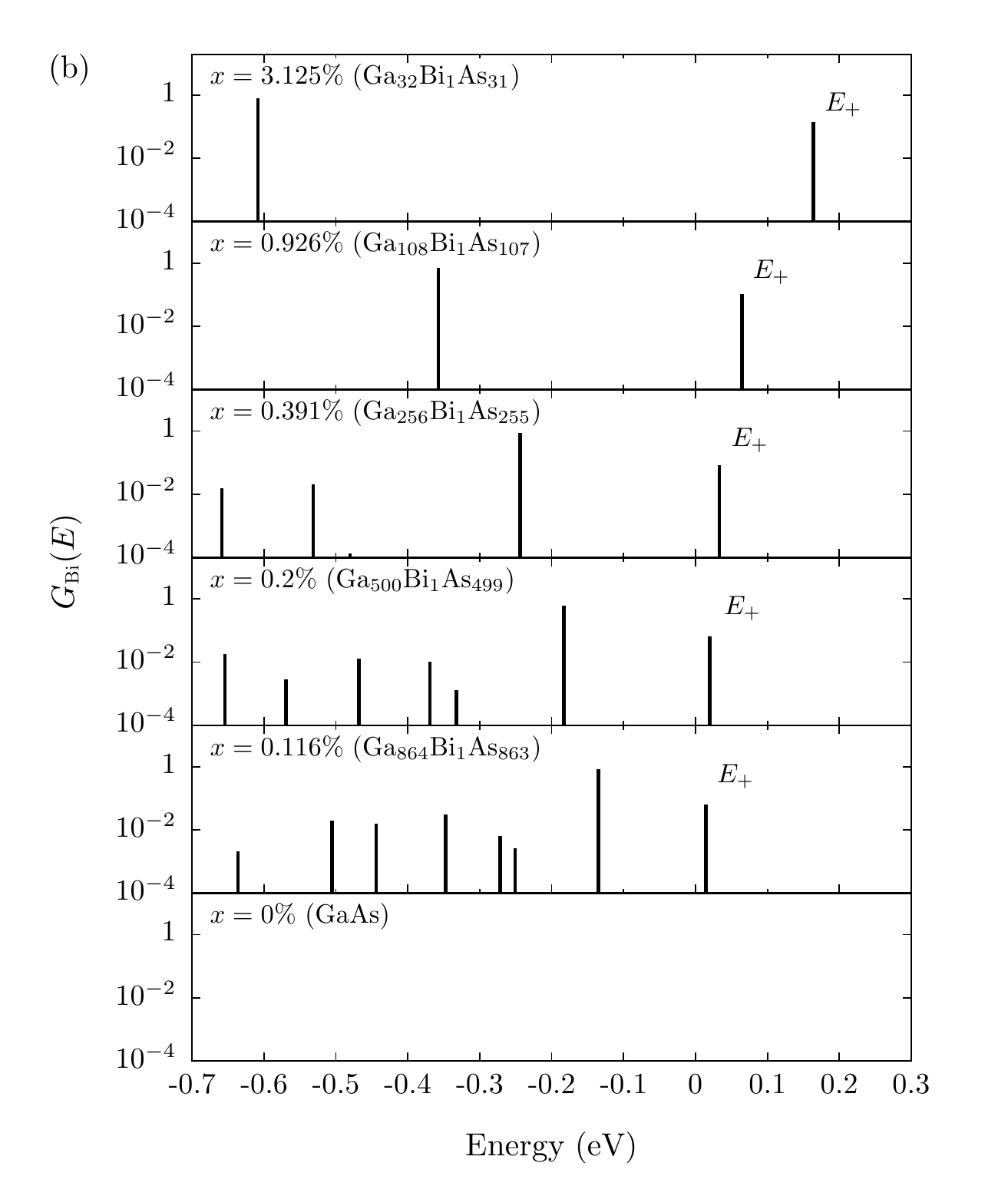} \label{fig:FGC_Bi_ordered_GaBiAs} }
	\caption{(a) Calculated $G_{\Gamma} (E)$ spectrum based on GaAs LH + HH states, and (b) calculated spectrum of fractional Bi localised state character, $G_{\textrm{Bi}} (E)$, for a series of ordered Ga$_{M}$Bi$_{1}$As$_{M-1}$ supercells. In each case the zero of energy is taken at the GaAs VBE and indicated by a vertical dashed line; note that $G_{\Gamma} (E)$ and $G_{\textrm{Bi}} (E)$ are plotted on a log scale, to highlight Ga$_{M}$Bi$_{1}$As$_{M-1}$ valence states with (small) non-zero $\Gamma$ and/or Bi character. The height of the bars in this figure indicates the fraction of each state at that energy which has (a) host matrix $\Gamma$ character, and (b) localised Bi state character.}
\end{figure*}

% Equation 3

\begin{equation}
	\vert \psi_{\scalebox{0.6}{\text{Bi}},i} \rangle = \dfrac{\vert \psi_{v,i}^{(1)} \rangle - \sum \limits_{n = 1}^{4} \vert \psi_{v,n}^{(0)} \rangle \langle \psi_{v,n}^{(0)} \vert \psi_{v,i}^{(1)} \rangle}{ \sqrt{ 1 - \sum \limits_{n = 1}^{4} \vert \langle \psi_{v,n}^{(0)} \vert \psi_{v,i}^{(1)} \rangle \vert^{2} }}, \; \; \; i = 1, \dots, 4
	\label{eq:Bi_state}
\end{equation}

\noindent
where $\vert \psi_{v,n}^{(0)} \rangle$ and $\vert \psi_{v,i}^{(1)} \rangle$ denote the VBE wave functions of the host and Bi-containing supercells, respectively, and the sums run over the four-fold degenerate states of the host matrix VBE. Using the full TB Hamiltonian for the Bi-containing supercell $\widehat{H}$ we can calculate the energy of the Bi-related impurity states and the strength of their interaction with the host matrix VBE as:

\begin{eqnarray}
	E_{\scalebox{0.6}{\textrm{Bi}},i} &=& \langle \psi_{\scalebox{0.6}{\textrm{Bi}},i} \vert \widehat{H} \vert \psi_{\scalebox{0.6}{\textrm{Bi}},i} \rangle \label{eq:E_Bi} \\
	V_{\scalebox{0.6}{\textrm{Bi}},i} &=& \langle \psi_{\scalebox{0.6}{\textrm{Bi}},i} \vert \widehat{H} \vert \psi_{v,0,i}                         \rangle \label{eq:V_Bi}
\end{eqnarray}

\noindent
where

\begin{equation}
	\vert \psi_{v,0,i} \rangle = \sum_{n = 1}^{4} \vert \psi_{v,n}^{(0)} \rangle \langle \psi_{v,n}^{(0)} \vert \psi_{v,i}^{(1)} \rangle
	\label{eq:host_VBE}
\end{equation}

\noindent
is the host matrix VBE state with which $\vert \psi_{\scalebox{0.6}{\textrm{Bi}},i} \rangle$ interacts.

By constructing the Bi-related impurity states in Ga$_{M}$Bi$_{1}$As$_{M-1}$ using Eq.~\eqref{eq:Bi_state} we have previously shown in Ref.~\onlinecite{Usman_PRB_2011} that (i) they interact with the Ga$_{M}$As$_{M}$ VBE via an anticrossing interaction $V_{\scalebox{0.6}{\textrm{Bi}},i} = \beta_{\scalebox{0.6}{\textrm{Bi}},i} \sqrt{x}$, and (ii) that this interaction explains the strong bowing of the band gap and spin-orbit-splitting present in \GaBiAs.
Using Eq.~\eqref{eq:Bi_state} we determine that the state which mixes with each of the GaAs VBE states is a localised, four-fold degenerate Bi resonant impurity state similar in character to that found in \GaBiP \cite{Usman_PRB_2011}. We also find that with decreasing supercell size (increasing Bi composition) the strength of the interaction between the resonant state and the GaAs VBE increases, pushing the alloy VBE, $E_{+}$, upwards in energy and mixing more of the GaAs VBE states with the Bi-related impurity states.

This can clearly be seen in Fig.~\ref{fig:FGC_VBM_ordered_GaBiAs}, which plots the calculated $G_{\Gamma}(E)$ spectra obtained by projecting the Ga$_{M}$As$_{M}$ VBE wave functions onto the spectrum of a series of ordered Ga$_{M}$Bi$_{1}$As$_{M-1}$ supercells. In each supercell we see that the $\Gamma$ character associated with the GaAs VBE is distributed over several supercell valence levels, with the broadening of the calculated $G_{\Gamma} (E)$ increasing with supercell size. Note that we use a log scale on the $y$-axis in Figs.~\ref{fig:FGC_VBM_ordered_GaBiAs} and~\ref{fig:FGC_Bi_ordered_GaBiAs} in order to make clear the small $\Gamma$ (or Bi) character associated with some of the lower energy valence states.

Figure~\ref{fig:FGC_Bi_ordered_GaBiAs} shows the calculated $G_{\scalebox{0.6}{\textrm{Bi}}} (E)$ spectra for the same set of Ga$_{M}$Bi$_{1}$As$_{M-1}$ supercells as in Fig.~\ref{fig:FGC_VBM_ordered_GaBiAs}, calculated by projecting the Bi impurity states, $\vert \psi_{\scalebox{0.6}{\textrm{Bi}},i} \rangle$ of Eq.~\eqref{eq:Bi_state} onto the full spectrum of the Bi-containing supercell. We see that the Bi-related states are also spread over several supercell valence levels, reflecting that these localised states are resonant with the GaAs valence states \cite{Usman_PRB_2011,Zhang_PRB_2005}. We attribute this broadening of the Bi impurity states to the large density of supercell zone centre host valence states with which the \psiBi\, states are resonant, with the number of such states increasing with supercell size as more bands fold back to $\Gamma$.

Comparing Figs.~\ref{fig:FGC_VBM_ordered_GaBiAs} and~\ref{fig:FGC_Bi_ordered_GaBiAs} we see that the GaAs VBE states and the Bi-related impurity states intermix in the Bi-containing supercells, so that the alloy VBE is comprised of an admixture of the host VBE and the Bi resonant states, consistent with a BAC interaction between them.

In \GaNAs\, the presence of an upper N-related BAC feature has been observed for $y \lesssim 3$\%, beyond which composition it broadens and weakens \cite{Perkins_PRL_1999} as the resonance becomes degenerate with the large density of conduction band states in the $L$-valley \cite{Lindsay_PE_2004}. The observed distribution of $G_{\Gamma}(E)$ across several valence states in \GaBiAs\, is then analogous to the \GaNAs\, case for $y \gtrsim 3$\%. We therefore attribute the failure to observe a feature associated with the lower energy Bi-related BAC levels, $E_{-}$, in PR spectra of \GaBiAs\, at any composition to the broadening and delocalisation of the Bi impurity states by the large density of host valence states with which they are resonant.

By examining the character of the impurity state of Eq.~\eqref{eq:Bi_state} associated with an isolated Bi impurity in increasingly large supercells, we approach the dilute doping limit \cite{Usman_PRB_2011,Broderick_PSSB_2013}. Table~\ref{tab:BAC_parameters} shows the calculated values of the energy of the four-fold degenerate Bi-related resonant states \EBi\, and the BAC coupling parameter \betaBi\, in 4096-atom GaAs and GaP supercells containing a single substitutional Bi atom. Based on the trends observed in the TB calculations, we conclude that Bi forms a resonant impurity level approximately 180 meV below the GaAs VBE in the dilute doping limit.

In summary, our TB calculations have shown that the VBE in ordered crystals of \GaBiP\, and \GaBiAs\, is well described in terms of a BAC interaction between the host matrix VBE and resonant impurity states associated with substitutional Bi atoms. Additionally, our calculations indicate that the conduction and spin-split-off band edges vary linearly in energy with increasing Bi composition and hence can be understood in terms of conventional alloying effects, without the need to include the SO-related BAC interaction that was assumed by Alberi \emph{et al.} \cite{Alberi_PRB_2007} in the first BAC model for \GaBiAs.

Having established the general form of the Bi-related interactions, we turn next to show that a 12-band \kdotp\, model, including the four Bi-related impurity levels, can be used to obtain an accurate description of the dispersion of the lowest conduction and highest valence bands in \GaBiAs, in excellent agreement with the results of TB supercell calculations.

% Table 1

\begin{table}[tb]
	\caption{\label{tab:BAC_parameters} Calculated values of the energy of the Bi impurity state \protect\EBi\, and BAC coupling parameter \protect\betaBi\, for Bi incorporation in GaAs and GaP in the dilute doping limit. \protect\EBi\, is given relative to the VBE of the host binary in each case.}
	\begin{ruledtabular}
		\begin{tabular}{ccc}
			& GaAs & GaP \\
			\hline
			\EBi   \;(eV) & -0.183 & 0.122 \\
			\betaBi\;(eV) &  1.13  & 1.41  \\
		\end{tabular}
	\end{ruledtabular}
\end{table}

% Table 2

\begin{table}[tb]
	\caption{\label{tab:VCA_parameters} Calculated virtual crystal contributions to the variation of the band edge energies with Bi and N composition in ordered Ga$_{M}$Bi$_{1}$As$_{M-1}$ and Ga$_{M}$N$_{1}$As$_{M-1}$ supercells.}
	\begin{ruledtabular}
		\begin{tabular}{ccc}
			& \GaBiAs & \GaNAs \\
			\hline
			$\alpha$ (eV) & 2.82 & -1.51 \\
			$\kappa$ (eV) & 1.01 &  1.36 \\
			$\gamma$ (eV) & 0.55 & -1.53 \\
		\end{tabular}
	\end{ruledtabular}
\end{table}

%%%%%%%%%%%%%%%%%%%%%%%%%%%%%%%%%%%%%%%%%%%%%
%%%% Section 3: 12-band model for GaBiAs %%%%
%%%%%%%%%%%%%%%%%%%%%%%%%%%%%%%%%%%%%%%%%%%%%

\section{12-band \kdotp\, model for dilute bismide alloys}
\label{sec:12band}

%%%%%%%%%%%%%%%%%%%%%%%%%%%%%%%%%%%%%%%%%
%%%% Section 3.1: Ordered supercells %%%%
%%%%%%%%%%%%%%%%%%%%%%%%%%%%%%%%%%%%%%%%%

\subsection{Derivation of the 12-band model for ordered supercells}
\label{sec:derivation}

We start in this section by considering the GaAs host matrix TB band structure, which we use to parameterise the corresponding 8-band \kdotp\, Hamiltonian to give an accurate description of the host matrix band dispersion close to the energy gap \cite{Lindsay_PSSB_1999}. Then, guided by TB calculations on ordered Ga$_{M}$Bi$_{1}$As$_{M-1}$ supercells we extend to a 12-band \kdotp\, model by including BAC interactions, $V_{\scalebox{0.6}{\textrm{Bi}}}$, between the GaAs VBE and the 4-fold degenerate Bi-related resonant impurity states at energy \EBi.

We begin by diagonalising the GaAs Hamiltonian at $\Gamma$, from which we obtain the 8-band basis states $\vert u_{1} \rangle, \dots, \vert u_{8} \rangle$. We then use these states to construct an 8-band parameter set for GaAs directly from the full TB calculations. This gives an 8-band \kdotp\, model that can reproduce the full TB band structure of a Ga$_{M}$As$_{M}$ supercell close to the energy gap at the zone centre.

The derived set of GaAs parameters represents a typical TB fit, with the calculated value of the Kane interband momentum matrix element $P = 11.1$ eV \AA\, being close to 75\% of its experimentally determined value \cite{Lindsay_PSSB_1999,Vurgaftman_JAP_2001}. This value of $P$ leads to a value of 16.08 eV for the Kane parameter $E_{P}$, compared to a typical value of 28.8 eV \cite{Vurgaftman_JAP_2001}. The band gap \Eg\, ($= 1.519$ eV) and spin-orbit-splitting energy \DeltaSO\, ($= 0.352$ eV) are taken directly from the calculated eigenvalues of the TB Hamiltonian at $\Gamma$. The Luttinger parameters can be obtained by fitting directly to the TB LH/HH band dispersion along the  $\Delta$-direction in the Brillouin zone ($\gamma_{1} = 5.22$, $\gamma_{2} = 1.42$), and the HH band dispersion along the $\Lambda$-direction in the Brillouin zone ($\gamma_{3} = 2.01$). Finally, the conduction band effective mass $m_{c}^{*}$ ($= 0.129$) is obtained by fitting to the TB-calculated dispersion of the lowest conduction band.

We note that this parametrisation of the 8-band model for GaAs, while non-standard, does not affect the validity of the conclusions to be drawn below. The model is used only to compare directly the TB- and \kdotp-calculated band dispersions, and enables us to demonstrate that, by beginning with an appropriately parametrised \kdotp\, model for GaAs, we can account for the effects of Bi incorporation not only on \Eg\, and \DeltaSO\, but also on the band dispersion of \GaBiAs. In particular, the parametrisation we have obtained for the 8-band model of GaAs has no bearing on the results presented below for the variation of \Eg\, and \DeltaSO\, with Bi and N composition in \GaBiAs\, and \GaBiNAs\, since all terms in the corresponding \kdotp\, Hamiltonians containing the parameters $E_{P}$, $m_{c}^{*}$, $\gamma_{1}$, $\gamma_{2}$ and $\gamma_{3}$ vanish at the $\Gamma$-point in the Brillouin zone. In general, we recommend that the Bi-related parameters presented in Tables~\ref{tab:BAC_parameters} and~\ref{tab:VCA_parameters} be combined with a standard set of GaAs \kdotp\, parameters, from e.g. Ref.~\onlinecite{Vurgaftman_JAP_2001}, when using the 12-band Hamiltonian to model \GaBiAs-based heterostructures.

Using the basis states obtained from the TB calculations we evaluate the dependence of the band edge energies on Bi composition $x$ as $\langle u_{i} \vert \widehat{H}(x) \vert u_{i} \rangle$, $i = 1, \dots, 8$. We calculate that these virtual crystal contributions to the variation of the GaAs band edge energies with Bi composition $x$ are given by:

% Equations 4, 5 and 6

\begin{eqnarray}
	E_{\scalebox{0.6}{\textrm{CB}}} (x) &=& E_{\scalebox{0.6}{\textrm{CB}}}                                         \left( \textrm{GaAs} \right) - \alpha x \label{eq:CB_VCA} \\
	E_{\scalebox{0.6}{\textrm{HH}}} (x)  =  E_{\scalebox{0.6}{\textrm{LH}}} (x) &=& E_{\scalebox{0.6}{\textrm{VB}}} \left( \textrm{GaAs} \right) + \kappa x \label{eq:VB_VCA} \\
	E_{\scalebox{0.6}{\textrm{SO}}} (x) &=& E_{\scalebox{0.6}{\textrm{SO}}}                                         \left( \textrm{GaAs} \right) - \gamma x \label{eq:SO_VCA}
\end{eqnarray}

\noindent
where the conduction and spin-split-off band edges vary linearly with Bi composition and are well described by Eqs.~\eqref{eq:CB_VCA} --~\eqref{eq:SO_VCA}. The calculated variations of the band edge energies in \GaBiAs\, and \GaNAs\, (obtained in the latter case by replacing $x$ with $y$ in Eqs.~\eqref{eq:CB_VCA} --~\eqref{eq:SO_VCA}) are listed in Table~\ref{tab:VCA_parameters}.

Based on the analysis of Section~\ref{sec:BAC_evidence} we then account for the effect of Bi incorporation on the GaAs VBE via a BAC interaction with coupling strength \betaBi$\sqrt{x}$, between the GaAs VBE and the four-fold degenerate Bi-related resonant impurity levels at energy \EBi. We choose these states to have HH and LH symmetry at the $\Gamma$-point, which then interact solely with the GaAs HH and LH states, respectively. The inclusion of these four degenerate impurity states into the \kdotp\, basis gives a 12-band \kdotp\, Hamiltonian for \GaBiAs:

% Equation 7

\begin{widetext}
	\begin{equation}
	\left( \begin{array}{cccccccccccc}
		E_{\scalebox{0.6}{\textrm{CB}}} (x) & -\sqrt{3}T_{+} & \sqrt{2}U & -U & 0 & 0 & 0 & 0 & -T_{-} & -\sqrt{2}T_{-} & 0 & 0 \\
		& E_{\scalebox{0.6}{\textrm{HH}}}(x) & \sqrt{2}S & -S & V_{\scalebox{0.6}{\textrm{Bi}}} & 0 & 0 & 0 & -R & -\sqrt{2}R & 0 & 0 \\
		& & E_{\scalebox{0.6}{\textrm{LH}}}(x) & Q & 0 & V_{\scalebox{0.6}{\textrm{Bi}}} & T_{+}^{*} & R & 0 & \sqrt{3}S & 0 & 0 \\
		& & & E_{\scalebox{0.6}{\textrm{SO}}}(x) & 0 & 0 & \sqrt{2}T_{+}^{*} & \sqrt{2}R & -\sqrt{3}S & 0 & 0 & 0 \\
		& & & & E_{\scalebox{0.6}{\textrm{Bi}}} & 0 & 0 & 0 & 0 & 0 & 0 & 0 \\
		& & & & & E_{\scalebox{0.6}{\textrm{Bi}}} & 0 & 0 & 0 & 0 & 0 & 0 \\
		& & & & & & E_{\scalebox{0.6}{\textrm{CB}}}(x) & -\sqrt{3}T_{-} & \sqrt{2}U & -U & 0 & 0 \\
		& & & & & & & E_{\scalebox{0.6}{\textrm{HH}}}(x) & \sqrt{2}S^{*} & -S^{*} & V_{\scalebox{0.6}{\textrm{Bi}}} & 0 \\
		& & & & & & & & E_{\scalebox{0.6}{\textrm{LH}}}(x) & Q & 0 & V_{\scalebox{0.6}{\textrm{Bi}}} \\
		& & & & & & & & & E_{\scalebox{0.6}{\textrm{SO}}}(x) & 0 & 0 \\
		& & & & & & & & & & E_{\scalebox{0.6}{\textrm{Bi}}} & 0 \\
		& & & & & & & & & & & E_{\scalebox{0.6}{\textrm{Bi}}} \\ \end{array} \right)
		\begin{array}{c}
			           \vert u_{1} \rangle \\
			           \vert u_{2} \rangle \\
			           \vert u_{3} \rangle \\
			           \vert u_{4} \rangle \\
		\hspace{0.23cm} \vert u_{\scalebox{0.6}{\textrm{Bi},1}}^{\scalebox{0.6}{\textrm{HH}}} \rangle \\
		\hspace{0.23cm} \vert u_{\scalebox{0.6}{\textrm{Bi},1}}^{\scalebox{0.6}{\textrm{LH}}} \rangle \\
			           \vert u_{5} \rangle \\
			           \vert u_{6} \rangle \\
			           \vert u_{7} \rangle \\
			           \vert u_{8} \rangle \\
		\hspace{0.23cm} \vert u_{\scalebox{0.6}{\textrm{Bi},2}}^{\scalebox{0.6}{\textrm{HH}}} \rangle \\
		\hspace{0.23cm} \vert u_{\scalebox{0.6}{\textrm{Bi},2}}^{\scalebox{0.6}{\textrm{LH}}} \rangle \\
		\end{array}
	\label{eq:12band_Hamiltonian}
	\end{equation}
\end{widetext}

\noindent
where $V_{\scalebox{0.6}{\text{Bi}}} = \beta_{\scalebox{0.6}{\text{Bi}}} \sqrt{x}$. We work here within the framework of the 8-band \kdotp\, model of Ref.~\onlinecite{Meney_PRB_1994} in which the definitions of the host matrix basis states $\vert u_{1} \rangle, \dots, \vert u_{8} \rangle$ and of the \textbf{k}-dependent matrix elements $R$, $S$, $T_{\pm}$ and $U$ can be found. The matrix elements $E_{\scalebox{0.6}{\textrm{CB}}}$, $E_{\scalebox{0.6}{\textrm{HH}}}$, $E_{\scalebox{0.6}{\textrm{LH}}}$, $E_{\scalebox{0.6}{\textrm{SO}}}$ are modified from those found in Ref.~\onlinecite{Meney_PRB_1994} by the inclusion of the $x$-dependent virtual crystal terms of Eqs.~\eqref{eq:CB_VCA} --~\eqref{eq:SO_VCA}. The Bi impurity states are taken to be dispersionless and the Bi-related basis states are listed in the Appendix. Since the Hamiltonian is a Hermitian matrix we neglect to show the below diagonal entries; they can be obtained from the above diagonals by Hermitian conjugation.

Using Eqs.~\eqref{eq:Bi_state} --~\eqref{eq:host_VBE}, we calculate the values of \EBi\, and $\beta_{\scalebox{0.6}{\textrm{Bi}}}$ in a series of Ga$_{M}$Bi$_{1}$As$_{M-1}$ supercells. This, combined with an appropriate 8-band parameter set for GaAs as well as the Bi compositional dependence of the band edges given by Eqs.~\eqref{eq:CB_VCA} --~\eqref{eq:SO_VCA} and Table~\ref{tab:VCA_parameters}, fully parameterizes the 12-band Hamiltonian.

%%%%%%%%%%%%%%%%%%%%%%%%%%%%%%%%%%%%%%%%%
%%%% Section 3.2: Ordered supercells %%%%
%%%%%%%%%%%%%%%%%%%%%%%%%%%%%%%%%%%%%%%%%

\subsection{Comparison of tight-binding and \protect\kdotp\, band structures in ordered supercells}
\label{sec:ordered}

\begin{figure*}[htb]
	\hspace{-1.0cm} \subfigure{ \scalebox{0.80}{\includegraphics{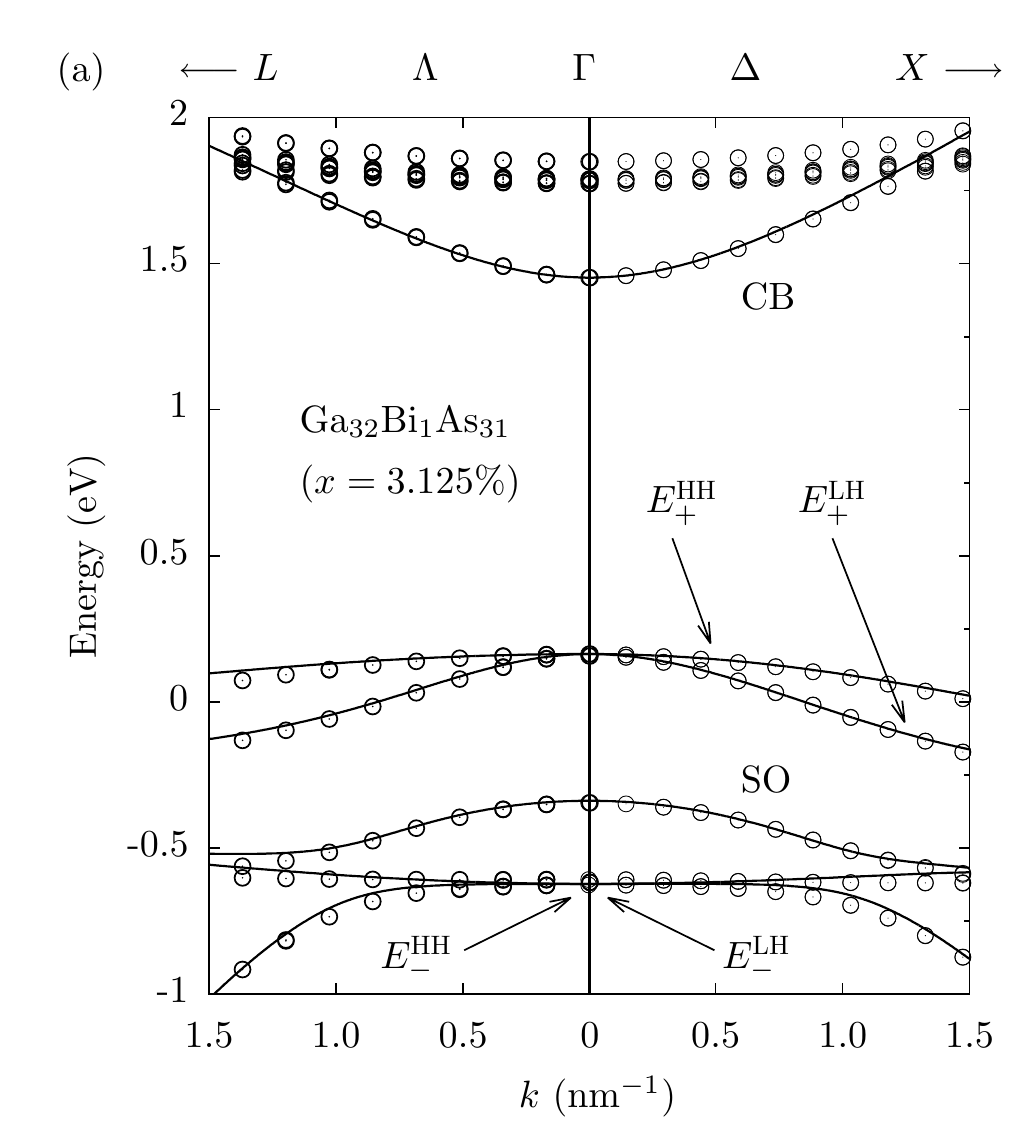}} \label{fig:Ga32Bi1As31_band_structure} }
	\hspace{-0.0cm} \subfigure{ \scalebox{0.80}{\includegraphics{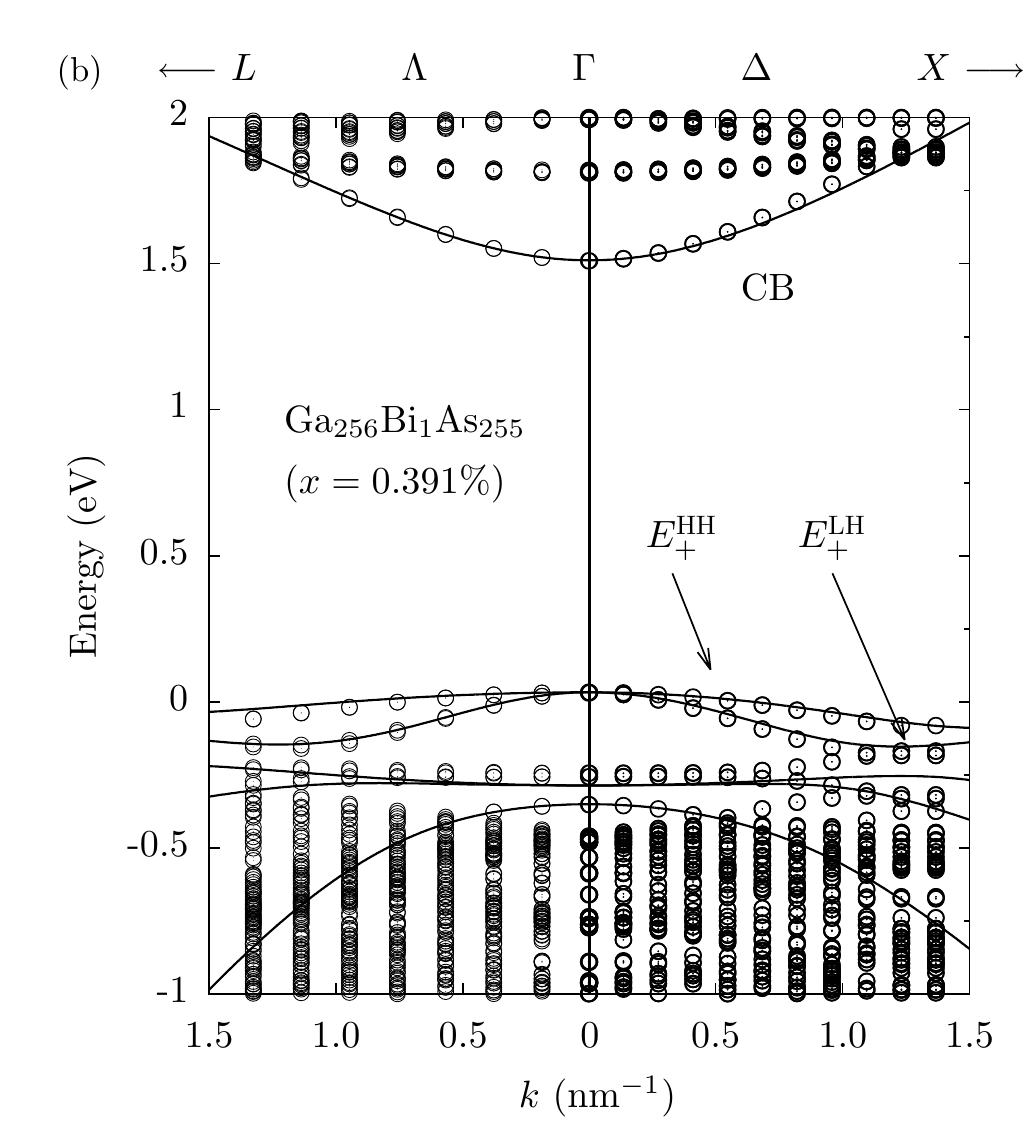}} \label{fig:Ga256Bi1As255_band_structure} }
	\caption{Calculated band dispersion of ordered, cubic (a) Ga$_{32}$Bi$_{1}$As$_{31}$ ($x = 3.125$\%), and (b) Ga$_{256}$Bi$_{1}$As$_{255}$ ($x = 0.391$\%) supercells, along the $\Lambda$- and $\Delta$-directions, close to the centre of the Brillouin zone, using the \protect\spss\, tight-binding (open circles) and 12-band \protect\kdotp\, (solid lines) Hamiltonians.}
\end{figure*}

Figures~\ref{fig:Ga32Bi1As31_band_structure} and~\ref{fig:Ga256Bi1As255_band_structure} compare the results of calculations of the Ga$_{32}$Bi$_{1}$As$_{31}$ ($x = 3.125$\%) and Ga$_{256}$Bi$_{1}$As$_{255}$ ($x = 0.391$\%) band dispersions along the $\Lambda$- and $\Delta$-directions in the vicinity of the zone centre using Eq.~\eqref{eq:12band_Hamiltonian} (solid lines) with those of full TB calculations (open circles). We see excellent agreement between the \kdotp\, and TB bands in both cases, and in particular we see that the presence of the BAC interaction is verified by the presence of TB bands that correspond to the lower energy Bi-related ($E_{-}$) BAC bands of the 12-band calculation.

% Figure 2

The calculations therefore confirm that the \GaBiAs\, energy gaps and band dispersion can be very well described through the inclusion of a BAC interaction between the GaAs VBE states and Bi-related impurity states below the band maximum.

We note however that the apparent direct correspondence between the Bi-impurity bands from the 12-band model and a specific set of TB bands in the 64-atom GaAs supercell occurs because the highest energy valence band folded back to the $\Gamma$-point in the 64-atom TB calculation lies approximately 1.4 eV below the VBE. There is a low density of supercell valence states with which the Bi impurity state is resonant. The Bi resonant states therefore remain unbroadened, contrary to the case in larger supercells containing a larger density of valence states (due to increased band folding as the size of the supercell Brillouin zone decreases). The BAC interaction is therefore clearly evident both for the upper and lower valence bands in Fig.~\ref{fig:Ga32Bi1As31_band_structure}.

In Refs.~\onlinecite{Usman_PRB_2011} and~\onlinecite{Broderick_PSSB_2013} we demonstrated that the density of host valence states plays a key role in determining \psiBi, resulting in a strong dependence of \EBi\, on supercell size. Specifically, the calculated value of \EBi\, shifts upwards in energy with increasing supercell size, as more GaAs valence bands fold back to $\Gamma$ and become available to construct \psiBi. We demonstrated that this trend stabilizes only as we move to large ($\gtrsim 2000$-atom) supercells, meaning that in order to calculate bulk \GaBiAs\, properties large supercell calculations were required.

It is for this reason that the Bi impurity state lies below the GaAs SO band in a 64-atom calculation but moves upward in energy as the supercell size is increased \cite{Broderick_PSSB_2013}, and it is also for this reason that the predicted values of \EBi\, and \betaBi\, are derived from supercell calculations containing $> 2000$ atoms, in which we approach the dilute doping limit.

The demonstration in Fig.~\ref{fig:Ga32Bi1As31_band_structure} that the BAC model can be used to describe the band dispersion of Ga$_{M}$Bi$_{1}$As$_{M-1}$ is contrary to the conclusion of a recent investigation of such supercells using DFT calculations \cite{Deng_PRB_2010}. From Eq.~\eqref{eq:Bi_state}, one would expect any lower valence state involved in the BAC interaction to show evidence of localisation about the Bi atom. The DFT calculations of Ref.~\onlinecite{Deng_PRB_2010} found that the second valence state in a 16-atom Ga$_{8}$Bi$_{1}$As$_{7}$ supercell showed no evidence of localisation about the Bi atom. We note however that this behavior is entirely consistent with the results presented in Fig.~\ref{fig:Ga32Bi1As31_band_structure}, where the Bi-related states have been pushed below the spin-split-off band \cite{Usman_PRB_2011,Broderick_PSSB_2013}, so that the second valence state in this case is then the delocalised spin-split-off band as opposed to a localised Bi-derived state. Further calculations that we have undertaken using our TB model show that the same situation also holds for a 16-atom supercell. We propose that the delocalised state observed in Ref.~\onlinecite{Deng_PRB_2010} is therefore most likely the spin-split-off band, and hence that the results presented there do not contradict the BAC model, as had previously been concluded.

Although the 12-band \kdotp\, model provides a good description of the band structure of \GaBiAs\, close to the energy gap, as demonstrated in Fig.~\ref{fig:Ga32Bi1As31_band_structure}, several points regarding the validity and interpretation of the 12-band model should be noted. In Ref.~\onlinecite{Lindsay_SSE_2003} it was demonstrated using a Green's function approach that the N-related impurity band in the bulk BAC Hamiltonian for \GaNAs\, should not be considered to represent a specific impurity band in the actual material band dispersion.  Rather, it represents a weighted average of all impurity-related states interacting with the band edge of the host matrix, reflecting the distribution of states, observed for instance in $G_{\Gamma} (E)$ and $G_{\scalebox{0.6}{\textrm{Bi}}} (E)$ in Figs.~\ref{fig:FGC_VBM_ordered_GaBiAs} and~\ref{fig:FGC_Bi_ordered_GaBiAs}. In the dilute nitride case the validity of the BAC model then followed from its ability to describe to a high degree of accuracy the modified dispersion of the lowest conduction band \cite{Lindsay_SSE_2003,Lindsay_SSC_2001}.

Similar behavior can be expected for \GaBiAs; as the host density of states increases (due to the folding of bands to $\Gamma$ as the size of the supercell is increased) the state associated with an isolated Bi atom becomes delocalised, with its associated character spreading over several supercell levels, which can be seen in Figs.~\ref{fig:FGC_VBM_ordered_GaBiAs} and~\ref{fig:FGC_Bi_ordered_GaBiAs}.

This is evident in Fig.~\ref{fig:Ga256Bi1As255_band_structure}, which compares the band structure of an ordered, cubic 512-atom Ga$_{256}$Bi$_{1}$As$_{255}$ supercell calculated using the \spss\, TB and 12-band \kdotp\, Hamiltonians. We see that the 12-band model provides an accurate description of the band edge energies and dispersions of the lowest-conduction, LH, HH and SO bands. However, in this calculation, the increased density of zone centre host valence states compared to the 64-atom case results in a broadening of the $\Gamma$ character associated with the lower BAC related bands, so that their associated $\Gamma$ character is spread over several Ga$_{256}$Bi$_{1}$As$_{255}$ valence levels, with lower valence states starting to accumulate more Bi character than in the smaller supercell calculations (cf. Fig.~\ref{fig:FGC_Bi_ordered_GaBiAs}).  This accounts for the small difference between the \kdotp\, and TB $\Gamma$ energies near $-0.3$ eV in the 512-atom calculation, where the \kdotp\, state is a weighted average of several states, as shown in Fig.~\ref{fig:FGC_VBM_ordered_GaBiAs}, and therefore lies just below the TB state energy.

In summary, the BAC model provides a valid and accurate description of the modified valence band structure of ordered \GaBiAs\, crystals, but with the caveat that the lower-lying Bi impurity bands are not necessarily associated with specific material bands -- rather, they describe the net effect of the Bi-related impurity states that interact with the host matrix VBE.

In the next section it will be demonstrated that this interpretation remains valid in the case of a disordered \GaBiAs\, alloy, where the quality of the agreement between the TB calculated band dispersion and that calculated using the 12-band model decreases somewhat due to a loss of both short and long-range order in the crystal. Nevertheless, the validity of the BAC model is further supported by showing that the variation of \Eg\, and \DeltaSO\, with Bi composition obtained from the 12-band model is in excellent agreement with full TB calculations and experimental measurements across the investigated composition range.

% Figure 3

\begin{figure*}[tb!]
	\hspace{-1.5cm} \subfigure{ \scalebox{0.85}{ \includegraphics{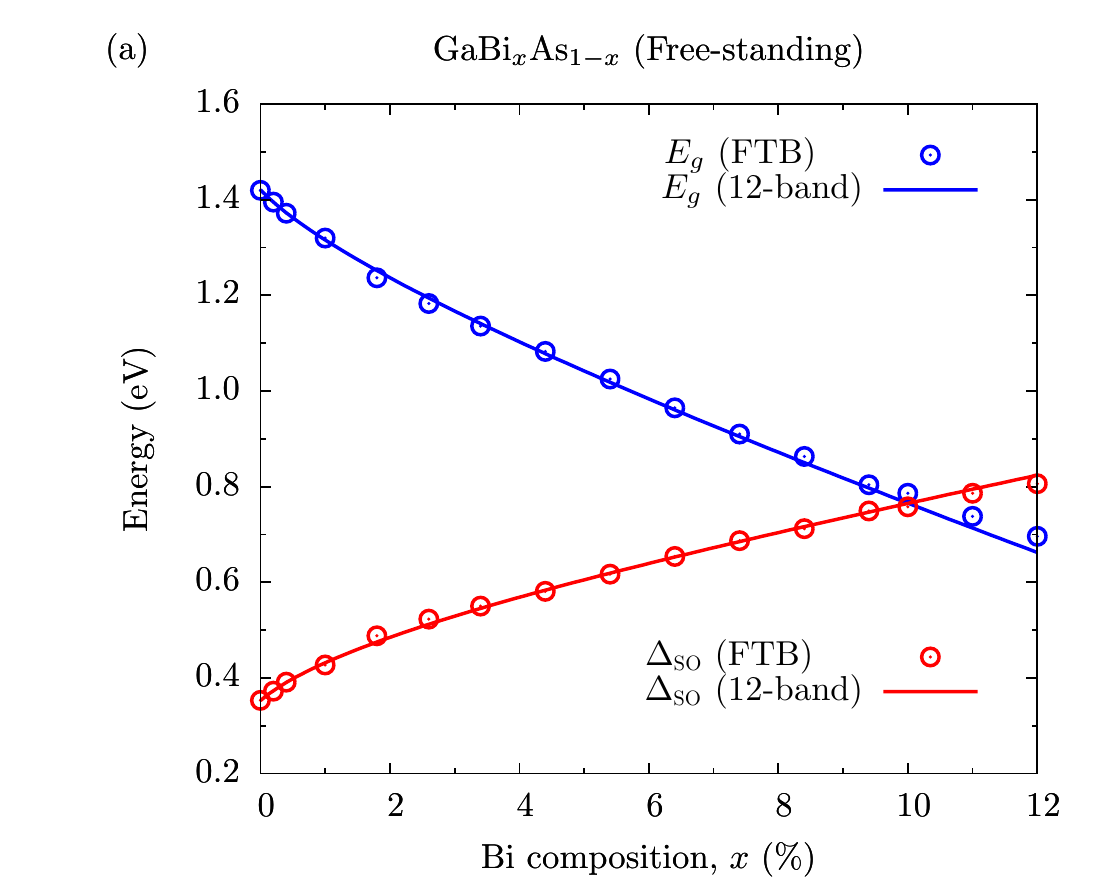}} \label{fig:Eg_SO_FTB_vs_kdotp}  }
	\hspace{-1.0cm} \subfigure{ \scalebox{0.85}{ \includegraphics{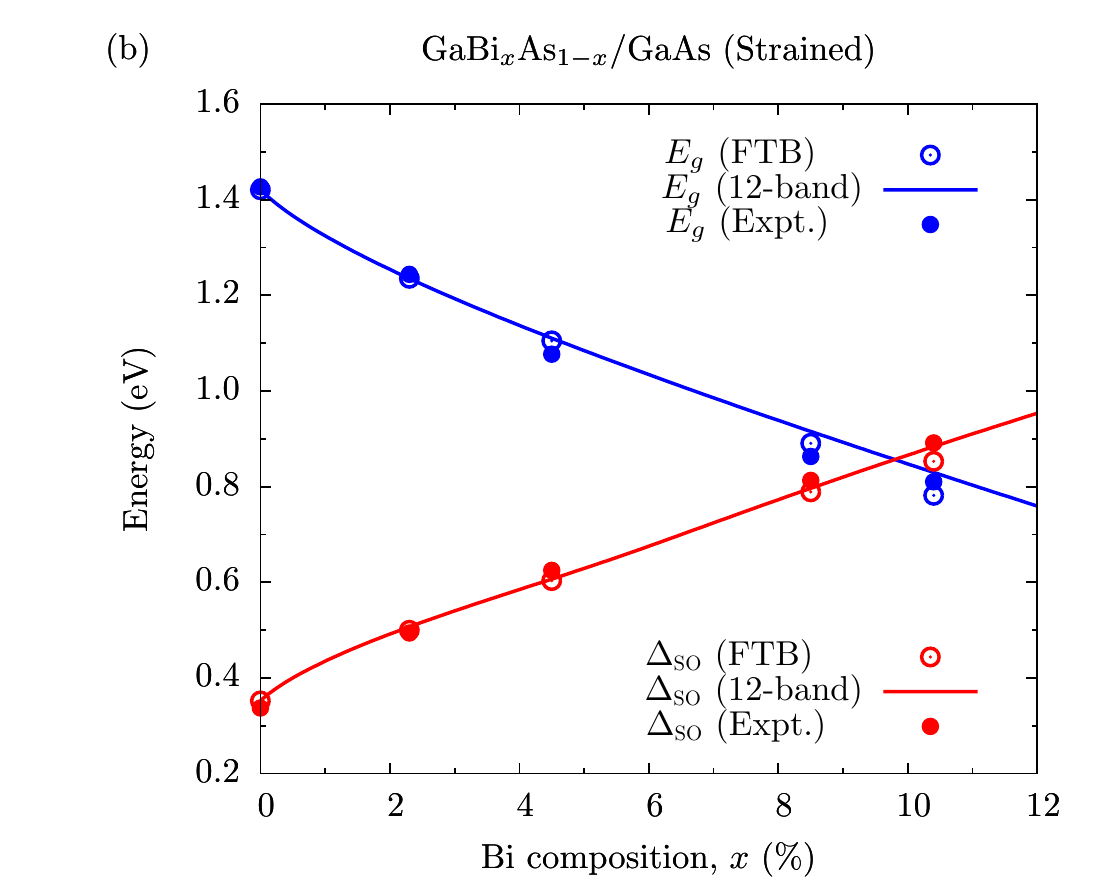}} \label{fig:Eg_SO_expt_vs_kdotp} }
	\caption{(a) Variation of the room temperature band gap (\protect\Eg) and spin-orbit-splitting (\protect\DeltaSO) as a function of Bi composition ($x$) in free-standing \protect\GaBiAs, calculated using the 12-band \kdotp\, model (blue and red solid lines) and full tight-binding (TB) calculations \cite{Usman_PRB_2011} on large, disordered supercells (blue and red open circles). (b) Variation of \protect\Eg\, and \protect\DeltaSO\, as a function of $x$ in biaxially strained \protect\GaBiAs/GaAs, calculated using the 12-band \kdotp\, model (blue and red solid lines) and full TB calculations \cite{Usman_PRB_2013} on large, disordered supercells (blue and red open circles) compared to photo-modulated reflectance measurements\cite{Batool_JAP_2012,Usman_PRB_2013} of \protect\Eg\, and \protect\DeltaSO\, as a function of $x$ on a series of \GaBiAs\, epilayers grown pseudomorphically on GaAs (blue and red closed circles).}
\end{figure*}

%%%%%%%%%%%%%%%%%%%%%%%%%%%%%%%%%%%%%%%%%%%%
%%%% Section 3.1: Disordered supercells %%%%
%%%%%%%%%%%%%%%%%%%%%%%%%%%%%%%%%%%%%%%%%%%%

\subsection{Application of the 12-band model to disordered supercells}
\label{sec:disordered}

The inclusion of multiple bismuth atoms to form a disordered Ga$_M$Bi$_L$As$_{M-L}$ supercell (containing $L$ substitutional bismuth atoms) can introduce Bi pairs (where a single gallium atom has two bismuth nearest-neighbours) and higher order clusters. Such disorder lowers the symmetry and lifts the degeneracy of the associated resonant states, $\vert \psi_{\scalebox{0.6}{\textrm{Bi}}, i} \rangle$. As we move towards a disordered alloy the GaAs VBE interacts with a distribution of impurity states related to isolated Bi atoms and Bi pairs and clusters, similar to those observed for N atoms in \GaNAs\cite{Lindsay_PRL_2004} and \GaNP \cite{Harris_JPCM_2008}. As predicted by the BAC model the alloy VBE shifts upward in energy and experiences an overall reduction in GaAs $\Gamma$ character with increasing Bi composition.

As the Bi composition increases in \GaBiAs\, the resulting disorder and formation of Bi pairs and clusters gives rise, in addition to the resonant states associated with isolated Bi atoms, to further impurity states which predominantly lie close in energy to the unperturbed GaAs VBE. These states then experience strong broadening due to the large density of GaAs valence states with which they are resonant, the result of which is a distribution of the GaAs VBE $\Gamma$ character related to Bi pair and cluster states over a large number of alloy valence states, due to the strong hybridisation between these states and those of the GaAs VBE \cite{Usman_PRB_2011}.

In order to demonstrate that the 12-band \kdotp\, model presented in Section~\ref{sec:derivation} is applicable to the realistic case of a disordered \GaBiAs\, alloy we use it here to investigate disordered supercells, for which one might expect the BAC model to break down. We firstly calculate \Eg\, and \DeltaSO\, in a series of large (4096-atom) \GaBiAs\, supercells in which Bi atoms are distributed in a statistically random way, leading to the formation of an increasing number of Bi pair and cluster states with increasing Bi composition $x$.

Figure~\ref{fig:Eg_SO_FTB_vs_kdotp} compares the room temperature band gap and spin-orbit-splitting energies calculated from TB calculations on large, disordered supercells \cite{Usman_PRB_2011} with those calculated using the 12-band model. We see that the two sets of calculations are in excellent agreement, confirming that the BAC model is capable of reproducing the variations of the band edge energies with Bi composition in an accurate manner over a large composition range.

Furthermore, we showed in Refs.~\onlinecite{Usman_PRB_2011} and~\onlinecite{Usman_PRB_2013} that the TB model we have developed for \GaBiAs\, is in very good agreement with a range of experimental measurements of \Eg\, and \DeltaSO\, in \GaBiAs, across the full composition range for which experimental data exists in the literature. In order to compare the 12-band model directly to experiment we have applied it to calculate the variation of \Eg\, and \DeltaSO\, with $x$ for \GaBiAs\, biaxially strained onto a GaAs substrate.

Figure~\ref{fig:Eg_SO_expt_vs_kdotp} shows the variation of \Eg\, and \DeltaSO\, in \GaBiAs\, biaxially strained onto a GaAs substrate, calculated using both the 12-band and TB models, and compared to PR measurements \cite{Batool_JAP_2012,Usman_PRB_2013} on a series of MBE-grown \GaBiAs/GaAs epilayers \cite{Lu_APL_2008,Lu_APL_2009}. The effects of biaxial strain were included in the 12-band calculations by using V\'{e}gard's law to interpolate between the GaAs and GaBi lattice and elastic constants in order to evaluate the components of the strain tensor and associated energy shifts to the band edges as a function of the Bi composition $x$. Biaxial strain effects were included in the TB calculations as outlined in Ref.~\onlinecite{Usman_PRB_2013}. We see from Fig.~\ref{fig:Eg_SO_expt_vs_kdotp} that the 12-band model is in good agreement with both the TB calculations and experiment across the investigated composition range, and that the 12-band model reproduces the important crossover to \Eg\, $<$\, \DeltaSO\, in the correct Bi composition range. We therefore conclude that the 12-band model presented here provides a realistic and accurate description of the main features of the \GaBiAs\, band structure.

% Figure 4

\begin{figure}[tb!]
	\scalebox{0.75}{\includegraphics{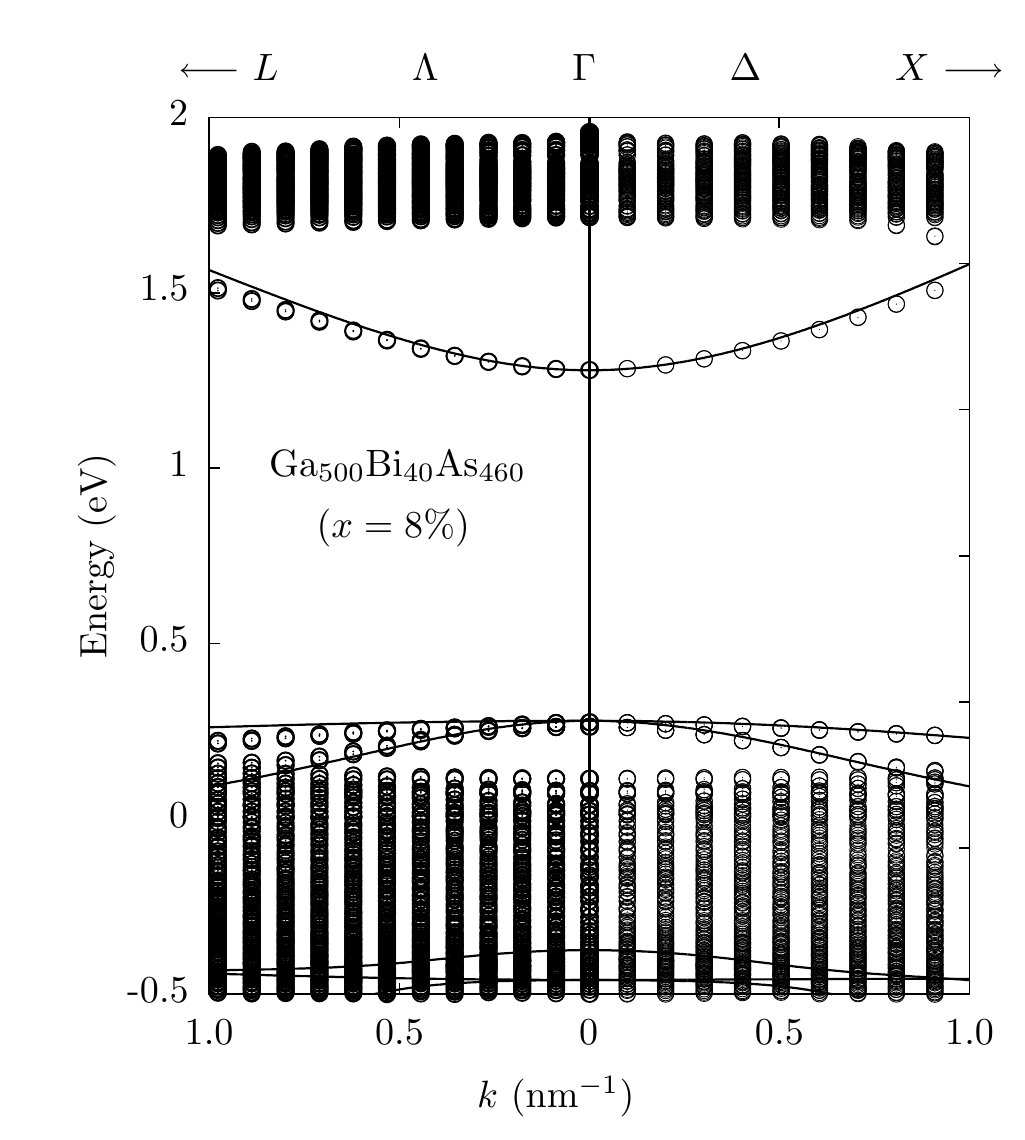}}
	\caption{Calculated band dispersion of a disordered Ga$_{500}$Bi$_{40}$As$_{460}$ ($x = 8$\%) supercell, along the $\Lambda$- and $\Delta$-directions, close to the centre of the Brillouin zone, using the \protect\spss\, tight-binding (open circles) and 12-band \protect\kdotp\, (solid lines) Hamiltonians.}
	\label{fig:Ga500Bi40As460_band_structure}
\end{figure}

Some limitations of the model as applied to disordered alloys should however be noted. In a disordered \GaBiAs\, alloy, the loss of long and short range order associated with the formation of Bi pairs and higher order clusters introduces localised distortions of the crystal lattice as well as giving rise to localised states lying close in energy to the GaAs VBE. Both of these effects perturb the band structure significantly, the first by reduction of symmetry, and the second by strong perturbation of the host valence band structure due to hybridisation with localised states.

While we have shown that the 12-band model including valence band-anticrossing provides an accurate description of the band edge energies at the $\Gamma$-point, we might expect that the quality of the description of the band dispersion away from $\Gamma$ should degrade somewhat compared to the ordered case of Section~\ref{sec:ordered} -- this indeed turns out to be the case. Figure~\ref{fig:Ga500Bi40As460_band_structure} compares the TB calculated band dispersion close to the zone centre of a disordered Ga$_{500}$Bi$_{40}$As$_{460}$ supercell (containing 8\% Bi) with that calculated using the 12-band model, in which it assumed that the alloy is ordered. As expected, the band edge energies at $\Gamma$ are well described by the 12-band model but the quality of the description of the band dispersion reduces more rapidly with increasing wave vector than in the ordered case of the previous subsection. Nevertheless, the 12-band model produces, even in the presence of significant alloy disorder, a satisfactory description of the band structure in the vicinity of the conduction and valence band edges at the $\Gamma$-point, which are the regions of interest for the calculation of the optoelectronic properties relevant to laser operation.

Having shown that the 12-band model introduced in Section~\ref{sec:derivation} is well-suited to describe the band structure of dilute bismide alloys close to the band edges and over a large composition range, we now turn our attention to dilute bismide-nitride alloys. We focus on the quaternary alloy \GaBiNAs \, and outline the extension of the 12-band model to a 14-band model which accounts explicitly for the effects of bismuth and nitrogen on the GaAs band structure.

%%%%%%%%%%%%%%%%%%%%%%%%%%%%%%%%%%%%%%%%%%%%%%
%%%% Section 5: 14-band model for GaBiNAs %%%%
%%%%%%%%%%%%%%%%%%%%%%%%%%%%%%%%%%%%%%%%%%%%%%

\section{Derivation of a 14-band \kdotp\, Hamiltonian for G\lowercase{a}B\lowercase{i}$_{x}$N$_{y}$A\lowercase{s}$_{1-x-y}$}
\label{sec:14_band_model}

In order to derive a 14-band Hamiltonian for \GaBiNAs\, from the TB model we first examine the effects of Bi and N when both are doped into GaAs in dilute quantities. We examine a series of ordered, cubic Ga$_{M}$Bi$_{1}$N$_{1}$As$_{M-2}$ supercells (in which it is ensured that the N-Bi seperation is as large as possible) using an \spss\, TB model for \GaBiNAs\, \cite{Usman_GaBiNAs_2013}.

% Figure 5

\begin{figure*}[tb!]
	\hspace{-1.0cm} \subfigure{ \scalebox{0.80}{\includegraphics{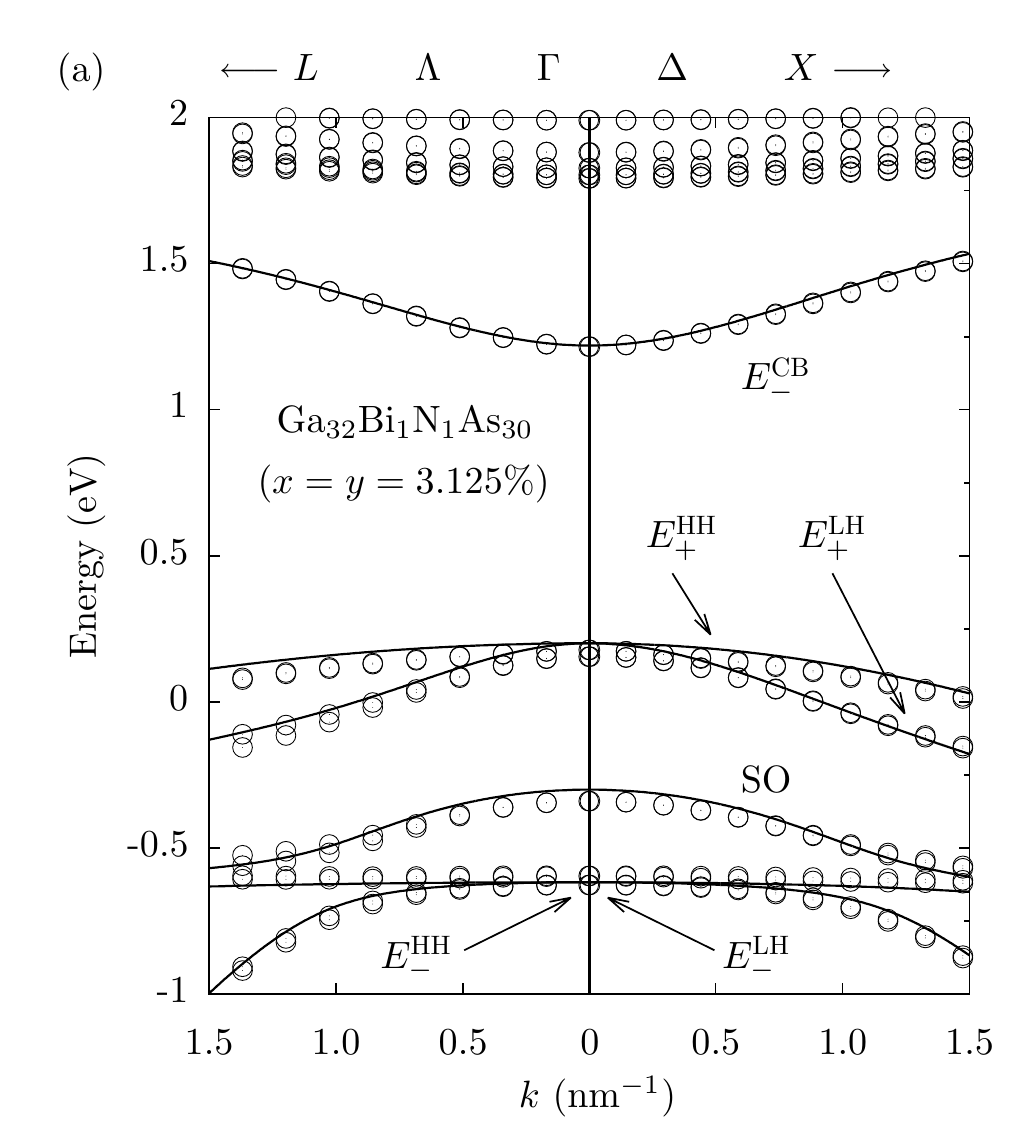}}   \label{fig:Ga32Bi1N1As30_band_structure} }
	\hspace{-0.0cm} \subfigure{ \scalebox{0.80}{\includegraphics{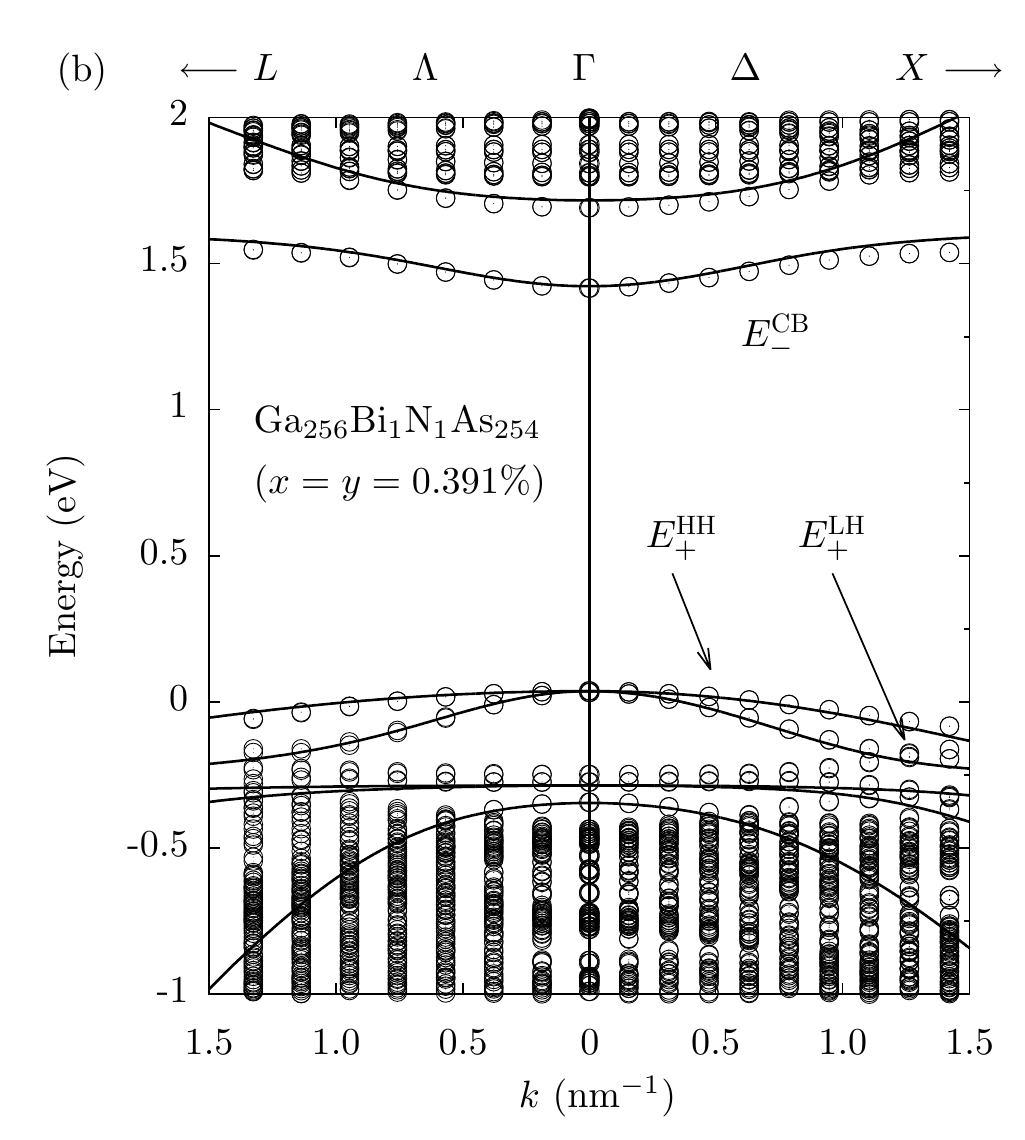}} \label{fig:Ga256Bi1N1As254_band_structure} }
	\caption{Calculated band dispersion of ordered, cubic (a) Ga$_{32}$Bi$_{1}$N$_{1}$As$_{30}$ ($x = y = 3.125$\%) and (b) Ga$_{256}$Bi$_{1}$N$_{1}$As$_{254}$ ($x = y = 0.391$\%) supercells, along the $\Lambda$- and $\Delta$-directions, close to the centre of the Brillouin zone, using the \spss\, (open circles) and 14-band \kdotp\, (solid lines) Hamiltonians. Note that the N-related impurity band in (a) is out of scale.}
\end{figure*}

To determine whether or not the effects of Bi and N on the GaAs electronic structure are independent of one another we construct the impurity wave functions associated with substitutional Bi and N atoms in these supercells and calculate their interaction by evaluating the matrix element of the full TB Hamiltonian for the Bi and N containing supercell between them: $V_{\scalebox{0.6}{\textrm{Bi,N}}} = \langle \psi_{\scalebox{0.6}{\textrm{N}}} \vert \widehat{H}_{\textrm{Ga}_{M}\textrm{Bi}_{1}\textrm{N}_{1}\textrm{As}_{M-2}} \vert \psi_{\scalebox{0.6}{\textrm{Bi}}} \rangle$.

This gives $V_{\scalebox{0.6}{\textrm{Bi,N}}} \lesssim 1$ $\mu$eV for all supercells considered, showing that the effects of isolated N and Bi atoms on the electronic structure are decoupled in the ordered case and hence that the electronic structure of ordered \GaBiNAs\, crystals admits a simple interpretation in terms of separate N- and Bi-related BAC interactions in the conduction and valence bands, respectively. Furthermore, in Ref.~\onlinecite{Usman_GaBiNAs_2013}, we show that the variations in the band edge energies of \GaBiNAs\, alloys can be described by combining the variations of the band edge energies in \GaBiAs\, and \GaNAs, even for large disordered supercells. This again emphasizes that Bi and N largely act independently of each other in \GaBiNAs\, and therefore their interactions with the host GaAs matrix can be well described by independent BAC interactions in the valence and conduction bands.

Guided by this insight, we modify the 12-band Hamiltonian of Eq.~\eqref{eq:12band_Hamiltonian} to account for the presence of N, by including two N-related states in the Hamiltonian, which have an anticrossing interaction with the conduction band \cite{Tomic_IEEEJSTQE_2003}. When parameterising the resulting 14-band Hamiltonian for a given \GaBiNAs\, supercell we treat the effects of Bi and N on the GaAs electronic structure as independent of one another. We begin with a Ga$_{M}$Bi$_{1}$N$_{1}$As$_{M-2}$ supercell and calculate the Bi (N) related BAC parameters in the equivalent N (Bi) free supercell. In this case the virtual crystal contributions to the variations of the band edge energies at the $\Gamma$-point as given by Eqs.~\eqref{eq:CB_VCA} --~\eqref{eq:SO_VCA} are modified as:

% Equations 8, 9 and 10

\begin{eqnarray*}
	E_{\scalebox{0.6}{\textrm{CB}}} (x,y) &=& E_{\scalebox{0.6}{\textrm{CB}}}                                           \left( \textrm{GaAs} \right) - \alpha_{\scalebox{0.6}{\textrm{Bi}}} x - \alpha_{\scalebox{0.6}{\textrm{N}}} y \\
	E_{\scalebox{0.6}{\textrm{HH}}} (x,y)  =  E_{\scalebox{0.6}{\textrm{LH}}} (x,y) &=& E_{\scalebox{0.6}{\textrm{VB}}} \left( \textrm{GaAs} \right) + \kappa_{\scalebox{0.6}{\textrm{Bi}}} x + \kappa_{\scalebox{0.6}{\textrm{N}}} y \\
	E_{\scalebox{0.6}{\textrm{SO}}} (x,y) &=& E_{\scalebox{0.6}{\textrm{SO}}}                                           \left( \textrm{GaAs} \right) - \gamma_{\scalebox{0.6}{\textrm{Bi}}} x - \gamma_{\scalebox{0.6}{\textrm{N}}} y
\end{eqnarray*}

\noindent
where the Bi- and N-related parameters $\alpha$, $\kappa$ and $\gamma$ are given in Table~\ref{tab:VCA_parameters}.

Figures~\ref{fig:Ga32Bi1N1As30_band_structure} and~\ref{fig:Ga256Bi1N1As254_band_structure} show the calculated band dispersion close to the zone centre of 64- and 512-atom Ga$_M$Bi$_1$N$_1$As$_{M-2}$ supercells using the \spss\, and 14-band \kdotp\, Hamiltonians, from which we see that the extension to a 14-band model accounts accurately for N and Bi co-alloying, reproducing accurately the band edge energies and modified band dispersion of the TB calculations.

\begin{figure}[tb!]
	\scalebox{0.75}{\includegraphics{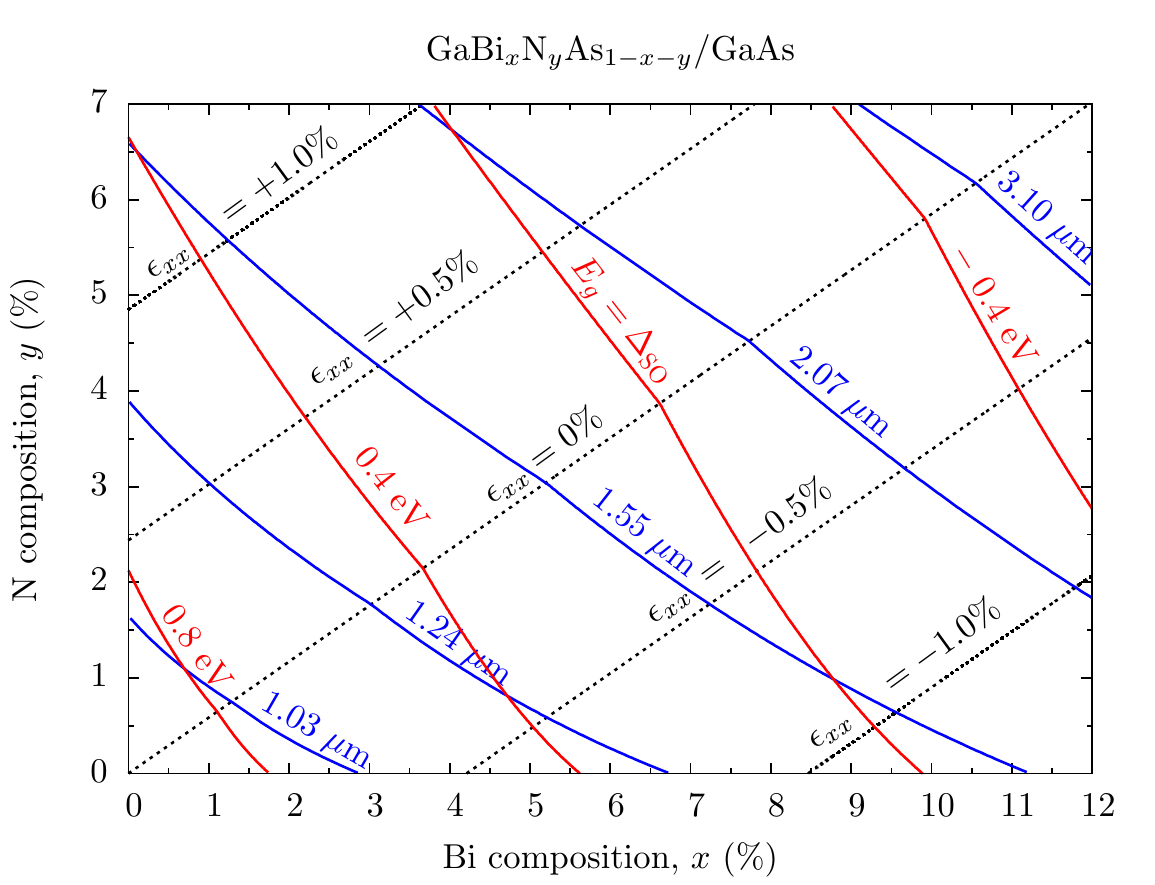}}
	\caption{Calculated variation of the band gap (\protect\Eg) and difference between the band gap and spin-orbit-splitting energy (\protect\Eg\, $-$\, \protect\DeltaSO) as a function of Bi and N composition ($x$ and $y$) for \protect\GaBiNAs/GaAs. Solid blue and red lines denote, respectively, paths in the composition space along which \protect\Eg\, and\, \protect\Eg\, $-$\, \protect\DeltaSO\, are constant. Dashed black lines denote paths in the composition space along which strain is constant. Alloys with compositions lying to the right of the \protect\Eg\, $=$\, \protect\DeltaSO\, contour are alloys in which \protect\DeltaSO\, $>$\, \protect\Eg\, and hence suppression of the dominant non-radiative CHSH Auger recombination pathway can be expected.}
	\label{fig:GaBiNAs_chart}
\end{figure}

Figure~\ref{fig:GaBiNAs_chart} shows the variation of the band gap (\Eg) and of the difference between the band gap and spin-orbit-splitting energy (\Eg\, $-$ \DeltaSO) for \GaBiNAs\, grown pseudomorphically on GaAs with $0 \leq x \leq 12$\% and $0 \leq y \leq 7$\%, for which the layers can be under compressive strain ($\epsilon_{xx} < 0$), strain-free ($\epsilon_{xx} = 0$), or under tensile strain ($\epsilon_{xx} > 0$) as the compositions are varied. In order to perform these calculations we have parametrised the 14-band Hamiltonian using the values of the GaAs room temperature band gap, spin-orbit-splitting energy, conduction band edge effective mass, Kane interband momentum matrix element, valence band Luttinger parameters, deformation potentials and lattice and elastic constants given in Ref.~\onlinecite{Vurgaftman_JAP_2001}. These 8-band \kdotp\, parameters for GaAs were then combined with the Bi and N related parameters listed in Tables~\ref{tab:BAC_parameters} and~\ref{tab:VCA_parameters}, where we took $E_{\scalebox{0.6}{\textrm{N}}} = 1.706$ eV and $\beta_{\scalebox{0.6}{\textrm{N}}} = 2.00$ eV \cite{Reilly_SST_2009}. The GaN and GaBi lattice and elastic constants were taken from Refs.~\onlinecite{Vurgaftman_JAP_2003} and~\onlinecite{Ferhat_PRB_2006}, with V\'{e}gard's law applied to calculate the \GaBiNAs\, lattice and elastic constants, and hence, the strain-induced shifts to the band edge energies.

Figure~\ref{fig:GaBiNAs_chart} shows that a very large wavelength range is accessible using \GaBiNAs\, at low strain on a GaAs substrate. The accessible wavelengths range from $\sim 1$ $\mu$m to wavelengths deep into the mid-infrared. Furthermore, the \Eg\, $=$ \DeltaSO\, contour in Fig.~\ref{fig:GaBiNAs_chart} indicates the combined Bi and N compositions beyond which a band gap energy lower than \DeltaSO\, can be achieved, which is essential for suppression of the non-radiative CHSH Auger recombination pathway. We see that by co-alloying Bi and N this \Eg\, $<$ \DeltaSO\, band structure condition can be obtained on GaAs at  1.55 $\mu$m and longer wavelengths, clearly demonstrating the potential of \GaBiAs\, and \GaBiNAs\, alloys for the design of highly efficient GaAs-based optoelectronic devices with reduced Auger losses across a wide wavelength range.

% Figure 6

%%%%%%%%%%%%%%%%%%%%%
%%%% Conclusions %%%%
%%%%%%%%%%%%%%%%%%%%%

\section{Conclusions}
\label{sec:conclusions}

Using a TB model which has previously been shown to be in good agreement with the main features of the \GaBiAs\, electronic structure deduced from experiment and from pseudopotential calculations, we have shown that the band structure of \GaBiAs\, can be well described by a band-anticrossing interaction between the extended GaAs valence band edge states and highly localised Bi-related resonant states, which lie below the GaAs valence band edge in energy.

Having shown in Ref.~\onlinecite{Usman_PRB_2011} that the observed strong bowing of the band gap and spin-orbit-splitting energies in \GaBiAs\, can be well understood in terms of this simple BAC model we have extended our analysis here to derive a 12-band \kdotp\, Hamiltonian for dilute bismide alloys, and demonstrated that it is capable of accurately reproducing key features of the band structure close to the band edges, for both ordered and disordered alloys as a function of Bi composition, $x$.

The 12-band model is in good agreement with full TB calculations of the variation of the (In)\GaBiAs\, band gap and spin-orbit-splitting energies with Bi composition, both for free-standing \GaBiAs\, and also with experimental measurements of the variation of the band gap and spin-orbit-splitting energies both of \GaBiAs\, epilayers grown on GaAs substrates, and of (In)\GaBiAs\, epilayers grown on InP substrates \cite{Broderick_PSSB_2013}. 

We have also shown that the effects of dilute co-alloying of N and Bi in GaAs are independent of one another in ordered supercells, a conclusion which has been confirmed by large supercell calculations on \GaBiNAs \cite{Broderick_ICTON_2011,Usman_GaBiNAs_2013}. This allowed us to introduce a 14-band \kdotp\, model for \GaBiNAs, which we showed to be in good agreement with full TB supercell calculations of the band structure.

We conclude that the 12 band and 14-band \kdotp\, models presented here provide simple and accurate descriptions of the band structure of dilute bismide and bismide-nitride alloys of GaAs in the vicinity of the band edges and hence will be of importance for the modeling and design of future (In)\GaBiAs\, and \GaBiNAs\, based optoelectronic devices.

%%%%%%%%%%%%%%%%%%%%%%%%%%
%%%% Acknowledgements %%%%
%%%%%%%%%%%%%%%%%%%%%%%%%%

\section*{Acknowledgements}

C. A. Broderick acknowledges financial support from the Irish Research Council under the EMBARK Initiative (RS/2010/2766). M. Usman and E. P. O'Reilly acknowledge financial support from the European Union Seventh Framework Programme (BIANCHO; FP7-257974).

%%%%%%%%%%%%%%%%%%%%%%%%%%%%%%%%
%%%% Appendix: Basis states %%%%
%%%%%%%%%%%%%%%%%%%%%%%%%%%%%%%%

\section*{Appendix}

The Bi-related basis states of the 12-band \kdotp\, Hamiltonian of Eq.~\eqref{eq:12band_Hamiltonian} are:

\begin{eqnarray*} 
	\vert u_{\scalebox{0.6}{\textrm{Bi,1}}}^{\scalebox{0.6}{\textrm{HH}}} \rangle &=& \left| \dfrac{3}{2}, + \dfrac{3}{2} \right\rangle = \frac{i}{\sqrt{2}} \vert x_{\scalebox{0.6}{\textrm{Bi}}} ; \uparrow \rangle - \frac{1}{\sqrt{2}} \vert y_{\scalebox{0.6}{\textrm{Bi}}} ; \uparrow \rangle \\
	\vert u_{\scalebox{0.6}{\textrm{Bi,2}}}^{\scalebox{0.6}{\textrm{HH}}} \rangle &=& \left| \dfrac{3}{2}, - \dfrac{3}{2} \right\rangle = - \frac{i}{\sqrt{2}} \vert x_{\scalebox{0.6}{\textrm{Bi}}} ; \downarrow \rangle - \frac{1}{\sqrt{2}} \vert y_{\scalebox{0.6}{\textrm{Bi}}} ; \downarrow \rangle \\
	\vert u_{\scalebox{0.6}{\textrm{Bi,1}}}^{\scalebox{0.6}{\textrm{LH}}} \rangle &=& \left| \dfrac{3}{2}, + \dfrac{1}{2} \right\rangle = \frac{i}{\sqrt{6}} \vert x_{\scalebox{0.6}{\textrm{Bi}}} ; \downarrow \rangle - \frac{1}{\sqrt{6}} \vert y_{\scalebox{0.6}{\textrm{Bi}}} ; \downarrow \rangle - \frac{i}{\sqrt{3}} \vert z_{\scalebox{0.6}{\textrm{Bi}}} ; \uparrow \rangle \\
	\vert u_{\scalebox{0.6}{\textrm{Bi,2}}}^{\scalebox{0.6}{\textrm{LH}}} \rangle &=& \left| \dfrac{3}{2}, - \dfrac{1}{2} \right\rangle = \frac{i}{\sqrt{6}} \vert x_{\scalebox{0.6}{\textrm{Bi}}} ; \uparrow \rangle + \frac{1}{\sqrt{6}} \vert y_{\scalebox{0.6}{\textrm{Bi}}} ; \uparrow \rangle + \frac{i}{\sqrt{3}} \vert z_{\scalebox{0.6}{\textrm{Bi}}} ; \downarrow \rangle
\end{eqnarray*}

%%%%%%%%%%%%%%%%%%%%
%%%% References %%%%
%%%%%%%%%%%%%%%%%%%%

% \bibliographystyle{apsrev}       % Using the Physical Review style for references
% \bibliography     {12and14band} % The bibliography is contained in the file: 12and14band.bib

\begin{thebibliography}{40}
\expandafter\ifx\csname natexlab\endcsname\relax\def\natexlab#1{#1}\fi
\expandafter\ifx\csname bibnamefont\endcsname\relax
  \def\bibnamefont#1{#1}\fi
\expandafter\ifx\csname bibfnamefont\endcsname\relax
  \def\bibfnamefont#1{#1}\fi
\expandafter\ifx\csname citenamefont\endcsname\relax
  \def\citenamefont#1{#1}\fi
\expandafter\ifx\csname url\endcsname\relax
  \def\url#1{\texttt{#1}}\fi
\expandafter\ifx\csname urlprefix\endcsname\relax\def\urlprefix{URL }\fi
\providecommand{\bibinfo}[2]{#2}
\providecommand{\eprint}[2][]{\url{#2}}

\bibitem[{\citenamefont{O'Reilly et~al.}(2009)\citenamefont{O'Reilly, Lindsay,
  Klar, Polimeni, and Capizzi}}]{Reilly_SST_2009}
\bibinfo{author}{\bibfnamefont{E.~P.} \bibnamefont{O'Reilly}},
  \bibinfo{author}{\bibfnamefont{A.}~\bibnamefont{Lindsay}},
  \bibinfo{author}{\bibfnamefont{P.~J.} \bibnamefont{Klar}},
  \bibinfo{author}{\bibfnamefont{A.}~\bibnamefont{Polimeni}}, \bibnamefont{and}
  \bibinfo{author}{\bibfnamefont{M.}~\bibnamefont{Capizzi}},
  \bibinfo{journal}{Semicond. Sci. Technol.} \textbf{\bibinfo{volume}{24}},
  \bibinfo{pages}{033001} (\bibinfo{year}{2009}).

\bibitem[{\citenamefont{Henini}(2005)}]{Henini}
\bibinfo{author}{\bibfnamefont{M.}~\bibnamefont{Henini}},
  \emph{\bibinfo{title}{Dilute Nitride Semiconductors}}
  (\bibinfo{publisher}{Elsevier}, \bibinfo{address}{Oxford},
  \bibinfo{year}{2005}).

\bibitem[{\citenamefont{Shan et~al.}(1999)\citenamefont{Shan, Walukiewicz,
  Ager, Haller, Geisz, Friedman, Olson, and Kurtz}}]{Shan_PRL_1999}
\bibinfo{author}{\bibfnamefont{W.}~\bibnamefont{Shan}},
  \bibinfo{author}{\bibfnamefont{W.}~\bibnamefont{Walukiewicz}},
  \bibinfo{author}{\bibfnamefont{J.~W.} \bibnamefont{Ager}},
  \bibinfo{author}{\bibfnamefont{E.~E.} \bibnamefont{Haller}},
  \bibinfo{author}{\bibfnamefont{J.~F.} \bibnamefont{Geisz}},
  \bibinfo{author}{\bibfnamefont{D.~J.} \bibnamefont{Friedman}},
  \bibinfo{author}{\bibfnamefont{J.~M.} \bibnamefont{Olson}}, \bibnamefont{and}
  \bibinfo{author}{\bibfnamefont{S.~R.} \bibnamefont{Kurtz}},
  \bibinfo{journal}{Phys. Rev. Lett.} \textbf{\bibinfo{volume}{82}},
  \bibinfo{pages}{1221} (\bibinfo{year}{1999}).

\bibitem[{\citenamefont{Alberi et~al.}(2007{\natexlab{a}})\citenamefont{Alberi,
  Wu, Walukiewicz, Yu, Dubon, Watkinsa, Wang, Liu, Cho, and
  Furdyna}}]{Alberi_PRB_2007}
\bibinfo{author}{\bibfnamefont{K.}~\bibnamefont{Alberi}},
  \bibinfo{author}{\bibfnamefont{J.}~\bibnamefont{Wu}},
  \bibinfo{author}{\bibfnamefont{W.}~\bibnamefont{Walukiewicz}},
  \bibinfo{author}{\bibfnamefont{K.~M.} \bibnamefont{Yu}},
  \bibinfo{author}{\bibfnamefont{O.~D.} \bibnamefont{Dubon}},
  \bibinfo{author}{\bibfnamefont{S.~P.} \bibnamefont{Watkinsa}},
  \bibinfo{author}{\bibfnamefont{C.~X.} \bibnamefont{Wang}},
  \bibinfo{author}{\bibfnamefont{X.}~\bibnamefont{Liu}},
  \bibinfo{author}{\bibfnamefont{Y.-J.} \bibnamefont{Cho}}, \bibnamefont{and}
  \bibinfo{author}{\bibfnamefont{J.}~\bibnamefont{Furdyna}},
  \bibinfo{journal}{Phys. Rev. B} \textbf{\bibinfo{volume}{75}},
  \bibinfo{pages}{045203} (\bibinfo{year}{2007}{\natexlab{a}}).

\bibitem[{\citenamefont{Tixier et~al.}(2005)\citenamefont{Tixier, Webster,
  Young, Tiedje, Francoeur, Mascarenhas, Wei, and
  Schiettekatte}}]{Tixier_APL_2005}
\bibinfo{author}{\bibfnamefont{S.}~\bibnamefont{Tixier}},
  \bibinfo{author}{\bibfnamefont{S.~E.} \bibnamefont{Webster}},
  \bibinfo{author}{\bibfnamefont{E.~C.} \bibnamefont{Young}},
  \bibinfo{author}{\bibfnamefont{T.}~\bibnamefont{Tiedje}},
  \bibinfo{author}{\bibfnamefont{S.}~\bibnamefont{Francoeur}},
  \bibinfo{author}{\bibfnamefont{A.}~\bibnamefont{Mascarenhas}},
  \bibinfo{author}{\bibfnamefont{P.}~\bibnamefont{Wei}}, \bibnamefont{and}
  \bibinfo{author}{\bibfnamefont{F.}~\bibnamefont{Schiettekatte}},
  \bibinfo{journal}{Appl. Phys. Lett.} \textbf{\bibinfo{volume}{86}},
  \bibinfo{pages}{112113} (\bibinfo{year}{2005}).

\bibitem[{\citenamefont{Batool et~al.}(2012)\citenamefont{Batool, Hild, Hosea,
  Lu, Tiedje, and Sweeney}}]{Batool_JAP_2012}
\bibinfo{author}{\bibfnamefont{Z.}~\bibnamefont{Batool}},
  \bibinfo{author}{\bibfnamefont{K.}~\bibnamefont{Hild}},
  \bibinfo{author}{\bibfnamefont{T.~J.~C.} \bibnamefont{Hosea}},
  \bibinfo{author}{\bibfnamefont{X.}~\bibnamefont{Lu}},
  \bibinfo{author}{\bibfnamefont{T.}~\bibnamefont{Tiedje}}, \bibnamefont{and}
  \bibinfo{author}{\bibfnamefont{S.~J.} \bibnamefont{Sweeney}},
  \bibinfo{journal}{J. Appl. Phys.} \textbf{\bibinfo{volume}{111}},
  \bibinfo{pages}{113108} (\bibinfo{year}{2012}).

\bibitem[{\citenamefont{Fluegel et~al.}(2006)\citenamefont{Fluegel, Francoeur,
  Mascarenhas, Tixier, Young, and Tiedje}}]{Fluegel_PRL_2006}
\bibinfo{author}{\bibfnamefont{B.}~\bibnamefont{Fluegel}},
  \bibinfo{author}{\bibfnamefont{S.}~\bibnamefont{Francoeur}},
  \bibinfo{author}{\bibfnamefont{A.}~\bibnamefont{Mascarenhas}},
  \bibinfo{author}{\bibfnamefont{S.}~\bibnamefont{Tixier}},
  \bibinfo{author}{\bibfnamefont{E.~C.} \bibnamefont{Young}}, \bibnamefont{and}
  \bibinfo{author}{\bibfnamefont{T.}~\bibnamefont{Tiedje}},
  \bibinfo{journal}{Phys. Rev. Lett.} \textbf{\bibinfo{volume}{97}},
  \bibinfo{pages}{067205} (\bibinfo{year}{2006}).

\bibitem[{\citenamefont{Usman et~al.}(2013)\citenamefont{Usman, Broderick,
  Batool, Hild, Hosea, Sweeney, and O'Reilly}}]{Usman_PRB_2013}
\bibinfo{author}{\bibfnamefont{M.}~\bibnamefont{Usman}},
  \bibinfo{author}{\bibfnamefont{C.~A.} \bibnamefont{Broderick}},
  \bibinfo{author}{\bibfnamefont{Z.}~\bibnamefont{Batool}},
  \bibinfo{author}{\bibfnamefont{K.}~\bibnamefont{Hild}},
  \bibinfo{author}{\bibfnamefont{T.~J.~C.} \bibnamefont{Hosea}},
  \bibinfo{author}{\bibfnamefont{S.~J.} \bibnamefont{Sweeney}},
  \bibnamefont{and} \bibinfo{author}{\bibfnamefont{E.~P.}
  \bibnamefont{O'Reilly}}, \bibinfo{journal}{Phys. Rev. B}
  \textbf{\bibinfo{volume}{87}}, \bibinfo{pages}{115104}
  (\bibinfo{year}{2013}).

\bibitem[{\citenamefont{Phillips et~al.}(1999)\citenamefont{Phillips, Sweeney,
  Adams, and Thijs}}]{Phillips_IEEEJSTQE_1999}
\bibinfo{author}{\bibfnamefont{A.~F.} \bibnamefont{Phillips}},
  \bibinfo{author}{\bibfnamefont{S.~J.} \bibnamefont{Sweeney}},
  \bibinfo{author}{\bibfnamefont{A.~R.} \bibnamefont{Adams}}, \bibnamefont{and}
  \bibinfo{author}{\bibfnamefont{P.~J.~A.} \bibnamefont{Thijs}},
  \bibinfo{journal}{IEEE J. Sel. Top. Quant. Electron.}
  \textbf{\bibinfo{volume}{5}}, \bibinfo{pages}{401} (\bibinfo{year}{1999}).

\bibitem[{\citenamefont{Sweeney et~al.}(2011)\citenamefont{Sweeney, Batool,
  Hild, Jin, and Hosea}}]{Sweeney_ICTON_2011}
\bibinfo{author}{\bibfnamefont{S.~J.} \bibnamefont{Sweeney}},
  \bibinfo{author}{\bibfnamefont{Z.}~\bibnamefont{Batool}},
  \bibinfo{author}{\bibfnamefont{K.}~\bibnamefont{Hild}},
  \bibinfo{author}{\bibfnamefont{S.~R.} \bibnamefont{Jin}}, \bibnamefont{and}
  \bibinfo{author}{\bibfnamefont{T.~J.~C.} \bibnamefont{Hosea}},
  \bibinfo{journal}{in Proceedings of the 13$^{\textrm{th}}$ International
  Conference on Transparent Optical Networks (ICTON), Stockholm}
  (\bibinfo{year}{2011}).

\bibitem[{\citenamefont{Alberi et~al.}(2007{\natexlab{b}})\citenamefont{Alberi,
  Dubon, Walukiewicz, Yu, Bertulis, and Krotkus}}]{Alberi_APL_2007}
\bibinfo{author}{\bibfnamefont{K.}~\bibnamefont{Alberi}},
  \bibinfo{author}{\bibfnamefont{O.~D.} \bibnamefont{Dubon}},
  \bibinfo{author}{\bibfnamefont{W.}~\bibnamefont{Walukiewicz}},
  \bibinfo{author}{\bibfnamefont{K.~M.} \bibnamefont{Yu}},
  \bibinfo{author}{\bibfnamefont{K.}~\bibnamefont{Bertulis}}, \bibnamefont{and}
  \bibinfo{author}{\bibfnamefont{A.}~\bibnamefont{Krotkus}},
  \bibinfo{journal}{Appl. Phys. Lett.} \textbf{\bibinfo{volume}{91}},
  \bibinfo{pages}{051909} (\bibinfo{year}{2007}{\natexlab{b}}).

\bibitem[{\citenamefont{Usman et~al.}(2011)\citenamefont{Usman, Broderick,
  Lindsay, and O'Reilly}}]{Usman_PRB_2011}
\bibinfo{author}{\bibfnamefont{M.}~\bibnamefont{Usman}},
  \bibinfo{author}{\bibfnamefont{C.~A.} \bibnamefont{Broderick}},
  \bibinfo{author}{\bibfnamefont{A.}~\bibnamefont{Lindsay}}, \bibnamefont{and}
  \bibinfo{author}{\bibfnamefont{E.~P.} \bibnamefont{O'Reilly}},
  \bibinfo{journal}{Phys. Rev. B} \textbf{\bibinfo{volume}{84}},
  \bibinfo{pages}{245202} (\bibinfo{year}{2011}).

\bibitem[{\citenamefont{Deng et~al.}(2010)\citenamefont{Deng, Li, Li, Peng,
  Xia, Wang, and Wei}}]{Deng_PRB_2010}
\bibinfo{author}{\bibfnamefont{H.-X.} \bibnamefont{Deng}},
  \bibinfo{author}{\bibfnamefont{J.}~\bibnamefont{Li}},
  \bibinfo{author}{\bibfnamefont{S.-S.} \bibnamefont{Li}},
  \bibinfo{author}{\bibfnamefont{H.}~\bibnamefont{Peng}},
  \bibinfo{author}{\bibfnamefont{J.-B.} \bibnamefont{Xia}},
  \bibinfo{author}{\bibfnamefont{L.-W.} \bibnamefont{Wang}}, \bibnamefont{and}
  \bibinfo{author}{\bibfnamefont{S.-H.} \bibnamefont{Wei}},
  \bibinfo{journal}{Phys. Rev. B} \textbf{\bibinfo{volume}{82}},
  \bibinfo{pages}{193204} (\bibinfo{year}{2010}).

\bibitem[{\citenamefont{Zhang et~al.}(2005)\citenamefont{Zhang, Mascarenhas,
  and Wang}}]{Zhang_PRB_2005}
\bibinfo{author}{\bibfnamefont{Y.}~\bibnamefont{Zhang}},
  \bibinfo{author}{\bibfnamefont{A.}~\bibnamefont{Mascarenhas}},
  \bibnamefont{and} \bibinfo{author}{\bibfnamefont{L.-W.} \bibnamefont{Wang}},
  \bibinfo{journal}{Phys. Rev. B} \textbf{\bibinfo{volume}{71}},
  \bibinfo{pages}{155201} (\bibinfo{year}{2005}).

\bibitem[{\citenamefont{Ludewig et~al.}(2013)\citenamefont{Ludewig, Knaub,
  Hossain, Reinhard, Marko, Jin, Hild, Chatterjee, Stolz, Sweeney
  et~al.}}]{Ludewig_APL_2013}
\bibinfo{author}{\bibfnamefont{P.}~\bibnamefont{Ludewig}},
  \bibinfo{author}{\bibfnamefont{N.}~\bibnamefont{Knaub}},
  \bibinfo{author}{\bibfnamefont{N.}~\bibnamefont{Hossain}},
  \bibinfo{author}{\bibfnamefont{S.}~\bibnamefont{Reinhard}},
  \bibinfo{author}{\bibfnamefont{I.~P.} \bibnamefont{Marko}},
  \bibinfo{author}{\bibfnamefont{S.~R.} \bibnamefont{Jin}},
  \bibinfo{author}{\bibfnamefont{K.}~\bibnamefont{Hild}},
  \bibinfo{author}{\bibfnamefont{S.}~\bibnamefont{Chatterjee}},
  \bibinfo{author}{\bibfnamefont{W.}~\bibnamefont{Stolz}},
  \bibinfo{author}{\bibfnamefont{S.~J.} \bibnamefont{Sweeney}},
  \bibnamefont{et~al.}, \bibinfo{journal}{Appl. Phys. Lett.}
  \textbf{\bibinfo{volume}{102}}, \bibinfo{pages}{242115}
  (\bibinfo{year}{2013}).

\bibitem[{\citenamefont{B\"{u}ckers et~al.}(2010)\citenamefont{B\"{u}ckers,
  K\"{u}hn, Schlichenmaier, Imhof, Thr\"{a}nhardt, Hader, Moloney, Rubel,
  Zhang, Ackemann et~al.}}]{Buckers_PSSB_2010}
\bibinfo{author}{\bibfnamefont{C.}~\bibnamefont{B\"{u}ckers}},
  \bibinfo{author}{\bibfnamefont{E.}~\bibnamefont{K\"{u}hn}},
  \bibinfo{author}{\bibfnamefont{C.}~\bibnamefont{Schlichenmaier}},
  \bibinfo{author}{\bibfnamefont{S.}~\bibnamefont{Imhof}},
  \bibinfo{author}{\bibfnamefont{A.}~\bibnamefont{Thr\"{a}nhardt}},
  \bibinfo{author}{\bibfnamefont{J.}~\bibnamefont{Hader}},
  \bibinfo{author}{\bibfnamefont{J.~V.} \bibnamefont{Moloney}},
  \bibinfo{author}{\bibfnamefont{O.}~\bibnamefont{Rubel}},
  \bibinfo{author}{\bibfnamefont{W.}~\bibnamefont{Zhang}},
  \bibinfo{author}{\bibfnamefont{T.}~\bibnamefont{Ackemann}},
  \bibnamefont{et~al.}, \bibinfo{journal}{Phys. Stat. Sol. B}
  \textbf{\bibinfo{volume}{247}}, \bibinfo{pages}{789} (\bibinfo{year}{2010}).

\bibitem[{\citenamefont{Chow et~al.}(1997)\citenamefont{Chow, Smowton, Blood,
  Girndt, Jahnke, and Koch}}]{Chow_APL_1997}
\bibinfo{author}{\bibfnamefont{W.~W.} \bibnamefont{Chow}},
  \bibinfo{author}{\bibfnamefont{P.~M.} \bibnamefont{Smowton}},
  \bibinfo{author}{\bibfnamefont{P.}~\bibnamefont{Blood}},
  \bibinfo{author}{\bibfnamefont{A.}~\bibnamefont{Girndt}},
  \bibinfo{author}{\bibfnamefont{F.}~\bibnamefont{Jahnke}}, \bibnamefont{and}
  \bibinfo{author}{\bibfnamefont{S.~W.} \bibnamefont{Koch}},
  \bibinfo{journal}{Appl. Phys. Lett.} \textbf{\bibinfo{volume}{71}},
  \bibinfo{pages}{157} (\bibinfo{year}{1997}).

\bibitem[{\citenamefont{Lermer et~al.}(2011)\citenamefont{Lermer,
  Gomez-Iglesias, Sabathil, Müller, Lutgen, Strauss, Pasenow, Hader, Moloney,
  Koch et~al.}}]{Lermer_APL_2011}
\bibinfo{author}{\bibfnamefont{T.}~\bibnamefont{Lermer}},
  \bibinfo{author}{\bibfnamefont{A.}~\bibnamefont{Gomez-Iglesias}},
  \bibinfo{author}{\bibfnamefont{M.}~\bibnamefont{Sabathil}},
  \bibinfo{author}{\bibfnamefont{J.}~\bibnamefont{Müller}},
  \bibinfo{author}{\bibfnamefont{S.}~\bibnamefont{Lutgen}},
  \bibinfo{author}{\bibfnamefont{U.}~\bibnamefont{Strauss}},
  \bibinfo{author}{\bibfnamefont{B.}~\bibnamefont{Pasenow}},
  \bibinfo{author}{\bibfnamefont{J.}~\bibnamefont{Hader}},
  \bibinfo{author}{\bibfnamefont{J.~V.} \bibnamefont{Moloney}},
  \bibinfo{author}{\bibfnamefont{S.~W.} \bibnamefont{Koch}},
  \bibnamefont{et~al.}, \bibinfo{journal}{Appl. Phys. Lett.}
  \textbf{\bibinfo{volume}{98}}, \bibinfo{pages}{021115}
  (\bibinfo{year}{2011}).

\bibitem[{\citenamefont{Thr\"{a}nhardt
  et~al.}(2005)\citenamefont{Thr\"{a}nhardt, Kuznetsova, Schlichenmaier, Koch,
  Shterengas, Belenky, Yeh, Mawst, Tansu, Hader et~al.}}]{Thranhardt_APL_2005}
\bibinfo{author}{\bibfnamefont{A.}~\bibnamefont{Thr\"{a}nhardt}},
  \bibinfo{author}{\bibfnamefont{I.}~\bibnamefont{Kuznetsova}},
  \bibinfo{author}{\bibfnamefont{C.}~\bibnamefont{Schlichenmaier}},
  \bibinfo{author}{\bibfnamefont{S.~W.} \bibnamefont{Koch}},
  \bibinfo{author}{\bibfnamefont{L.}~\bibnamefont{Shterengas}},
  \bibinfo{author}{\bibfnamefont{G.}~\bibnamefont{Belenky}},
  \bibinfo{author}{\bibfnamefont{J.-Y.} \bibnamefont{Yeh}},
  \bibinfo{author}{\bibfnamefont{L.~J.} \bibnamefont{Mawst}},
  \bibinfo{author}{\bibfnamefont{N.}~\bibnamefont{Tansu}},
  \bibinfo{author}{\bibfnamefont{J.}~\bibnamefont{Hader}},
  \bibnamefont{et~al.}, \bibinfo{journal}{Appl. Phys. Lett.}
  \textbf{\bibinfo{volume}{86}}, \bibinfo{pages}{021117}
  (\bibinfo{year}{2005}).

\bibitem[{\citenamefont{Broderick et~al.}(2012)\citenamefont{Broderick, Usman,
  Sweeney, and O'Reilly}}]{Broderick_SST_2012}
\bibinfo{author}{\bibfnamefont{C.~A.} \bibnamefont{Broderick}},
  \bibinfo{author}{\bibfnamefont{M.}~\bibnamefont{Usman}},
  \bibinfo{author}{\bibfnamefont{S.~J.} \bibnamefont{Sweeney}},
  \bibnamefont{and} \bibinfo{author}{\bibfnamefont{E.~P.}
  \bibnamefont{O'Reilly}}, \bibinfo{journal}{Semicond. Sci. Technol.}
  \textbf{\bibinfo{volume}{27}}, \bibinfo{pages}{094011}
  (\bibinfo{year}{2012}).

\bibitem[{\citenamefont{Sweeney and Jin}(2013)}]{Sweeney_JAP_2013}
\bibinfo{author}{\bibfnamefont{S.~J.} \bibnamefont{Sweeney}} \bibnamefont{and}
  \bibinfo{author}{\bibfnamefont{S.~R.} \bibnamefont{Jin}},
  \bibinfo{journal}{J. Appl. Phys.} \textbf{\bibinfo{volume}{113}},
  \bibinfo{pages}{043110} (\bibinfo{year}{2013}).

\bibitem[{\citenamefont{Janotti et~al.}(2002)\citenamefont{Janotti, Wei, and
  Zhang}}]{Janotti_PRB_2002}
\bibinfo{author}{\bibfnamefont{A.}~\bibnamefont{Janotti}},
  \bibinfo{author}{\bibfnamefont{S.-H.} \bibnamefont{Wei}}, \bibnamefont{and}
  \bibinfo{author}{\bibfnamefont{S.~B.} \bibnamefont{Zhang}},
  \bibinfo{journal}{Phys. Rev. B} \textbf{\bibinfo{volume}{65}},
  \bibinfo{pages}{115203} (\bibinfo{year}{2002}).

\bibitem[{\citenamefont{Broderick et~al.}(DOI:
  10.1109/ICTON.2011.5970828)\citenamefont{Broderick, Usman, and
  O'Reilly}}]{Broderick_ICTON_2011}
\bibinfo{author}{\bibfnamefont{C.~A.} \bibnamefont{Broderick}},
  \bibinfo{author}{\bibfnamefont{M.}~\bibnamefont{Usman}}, \bibnamefont{and}
  \bibinfo{author}{\bibfnamefont{E.~P.} \bibnamefont{O'Reilly}},
  \bibinfo{journal}{in Proceedings of the 13$^{\textrm{th}}$ International
  Conference on Transparent Optical Networks (ICTON), Stockholm 2011}
  (\bibinfo{year}{DOI: 10.1109/ICTON.2011.5970828}).

\bibitem[{\citenamefont{Usman et~al.}(in preparation, 2013)\citenamefont{Usman,
  Broderick, and O'Reilly}}]{Usman_GaBiNAs_2013}
\bibinfo{author}{\bibfnamefont{M.}~\bibnamefont{Usman}},
  \bibinfo{author}{\bibfnamefont{C.~A.} \bibnamefont{Broderick}},
  \bibnamefont{and} \bibinfo{author}{\bibfnamefont{E.~P.}
  \bibnamefont{O'Reilly}}, \bibinfo{journal}{An atomistic understanding of the
  electronic structure of GaAs-bismides-nitrides}  (\bibinfo{year}{in
  preparation, 2013}).

\bibitem[{\citenamefont{Trumbore et~al.}(1966)\citenamefont{Trumbore,
  Gershenzon, and Thomas}}]{Trumbore_APL_1964}
\bibinfo{author}{\bibfnamefont{F.~A.} \bibnamefont{Trumbore}},
  \bibinfo{author}{\bibfnamefont{M.}~\bibnamefont{Gershenzon}},
  \bibnamefont{and} \bibinfo{author}{\bibfnamefont{D.~G.}
  \bibnamefont{Thomas}}, \bibinfo{journal}{Appl. Phys. Lett.}
  \textbf{\bibinfo{volume}{9}}, \bibinfo{pages}{4} (\bibinfo{year}{1966}).

\bibitem[{\citenamefont{Perkins et~al.}(1999)\citenamefont{Perkins,
  Mascarenhas, Zhang, Geisz, Friedman, Olson, and Kurtz}}]{Perkins_PRL_1999}
\bibinfo{author}{\bibfnamefont{J.~D.} \bibnamefont{Perkins}},
  \bibinfo{author}{\bibfnamefont{A.}~\bibnamefont{Mascarenhas}},
  \bibinfo{author}{\bibfnamefont{Y.}~\bibnamefont{Zhang}},
  \bibinfo{author}{\bibfnamefont{J.~F.} \bibnamefont{Geisz}},
  \bibinfo{author}{\bibfnamefont{D.~J.} \bibnamefont{Friedman}},
  \bibinfo{author}{\bibfnamefont{J.~M.} \bibnamefont{Olson}}, \bibnamefont{and}
  \bibinfo{author}{\bibfnamefont{S.~R.} \bibnamefont{Kurtz}},
  \bibinfo{journal}{Phys. Rev. Lett.} \textbf{\bibinfo{volume}{82}},
  \bibinfo{pages}{3312} (\bibinfo{year}{1999}).

\bibitem[{\citenamefont{Lindsay and
  O'Reilly}(2004{\natexlab{a}})}]{Lindsay_PE_2004}
\bibinfo{author}{\bibfnamefont{A.}~\bibnamefont{Lindsay}} \bibnamefont{and}
  \bibinfo{author}{\bibfnamefont{E.~P.} \bibnamefont{O'Reilly}},
  \bibinfo{journal}{Physica E} \textbf{\bibinfo{volume}{21}},
  \bibinfo{pages}{901} (\bibinfo{year}{2004}{\natexlab{a}}).

\bibitem[{\citenamefont{Broderick et~al.}(2013)\citenamefont{Broderick, Usman,
  and O'Reilly}}]{Broderick_PSSB_2013}
\bibinfo{author}{\bibfnamefont{C.~A.} \bibnamefont{Broderick}},
  \bibinfo{author}{\bibfnamefont{M.}~\bibnamefont{Usman}}, \bibnamefont{and}
  \bibinfo{author}{\bibfnamefont{E.~P.} \bibnamefont{O'Reilly}},
  \bibinfo{journal}{Phys. Stat. Sol. B} \textbf{\bibinfo{volume}{250}},
  \bibinfo{pages}{773} (\bibinfo{year}{2013}).

\bibitem[{\citenamefont{Lindsay and O'Reilly}(1999)}]{Lindsay_PSSB_1999}
\bibinfo{author}{\bibfnamefont{A.}~\bibnamefont{Lindsay}} \bibnamefont{and}
  \bibinfo{author}{\bibfnamefont{E.~P.} \bibnamefont{O'Reilly}},
  \bibinfo{journal}{Phys. Stat. Sol. B} \textbf{\bibinfo{volume}{216}},
  \bibinfo{pages}{131} (\bibinfo{year}{1999}).

\bibitem[{\citenamefont{Vurgaftman et~al.}(2001)\citenamefont{Vurgaftman,
  Meyer, and Ram-Mohan}}]{Vurgaftman_JAP_2001}
\bibinfo{author}{\bibfnamefont{I.}~\bibnamefont{Vurgaftman}},
  \bibinfo{author}{\bibfnamefont{J.~R.} \bibnamefont{Meyer}}, \bibnamefont{and}
  \bibinfo{author}{\bibfnamefont{L.~R.} \bibnamefont{Ram-Mohan}},
  \bibinfo{journal}{J. Appl. Phys.} \textbf{\bibinfo{volume}{89}},
  \bibinfo{pages}{5815} (\bibinfo{year}{2001}).

\bibitem[{\citenamefont{Meney et~al.}(1994)\citenamefont{Meney, Gonul, and
  O'Reilly}}]{Meney_PRB_1994}
\bibinfo{author}{\bibfnamefont{A.~T.} \bibnamefont{Meney}},
  \bibinfo{author}{\bibfnamefont{B.}~\bibnamefont{Gonul}}, \bibnamefont{and}
  \bibinfo{author}{\bibfnamefont{E.~P.} \bibnamefont{O'Reilly}},
  \bibinfo{journal}{Phys. Rev. B} \textbf{\bibinfo{volume}{50}},
  \bibinfo{pages}{10893} (\bibinfo{year}{1994}).

\bibitem[{\citenamefont{Lindsay et~al.}(2003)\citenamefont{Lindsay, Tomi\'{c},
  and O'Reilly}}]{Lindsay_SSE_2003}
\bibinfo{author}{\bibfnamefont{A.}~\bibnamefont{Lindsay}},
  \bibinfo{author}{\bibfnamefont{S.}~\bibnamefont{Tomi\'{c}}},
  \bibnamefont{and} \bibinfo{author}{\bibfnamefont{E.~P.}
  \bibnamefont{O'Reilly}}, \bibinfo{journal}{Solid State Electron.}
  \textbf{\bibinfo{volume}{47}}, \bibinfo{pages}{443} (\bibinfo{year}{2003}).

\bibitem[{\citenamefont{Lindsay and O'Reilly}(2001)}]{Lindsay_SSC_2001}
\bibinfo{author}{\bibfnamefont{A.}~\bibnamefont{Lindsay}} \bibnamefont{and}
  \bibinfo{author}{\bibfnamefont{E.~P.} \bibnamefont{O'Reilly}},
  \bibinfo{journal}{Solid State Commun.} \textbf{\bibinfo{volume}{118}},
  \bibinfo{pages}{313} (\bibinfo{year}{2001}).

\bibitem[{\citenamefont{Lindsay and
  O'Reilly}(2004{\natexlab{b}})}]{Lindsay_PRL_2004}
\bibinfo{author}{\bibfnamefont{A.}~\bibnamefont{Lindsay}} \bibnamefont{and}
  \bibinfo{author}{\bibfnamefont{E.~P.} \bibnamefont{O'Reilly}},
  \bibinfo{journal}{Phys. Rev. Lett.} \textbf{\bibinfo{volume}{93}},
  \bibinfo{pages}{196402} (\bibinfo{year}{2004}{\natexlab{b}}).

\bibitem[{\citenamefont{Harris et~al.}(2008)\citenamefont{Harris, Lindsay, and
  O'Reilly}}]{Harris_JPCM_2008}
\bibinfo{author}{\bibfnamefont{C.}~\bibnamefont{Harris}},
  \bibinfo{author}{\bibfnamefont{A.}~\bibnamefont{Lindsay}}, \bibnamefont{and}
  \bibinfo{author}{\bibfnamefont{E.~P.} \bibnamefont{O'Reilly}},
  \bibinfo{journal}{J. Phys.: Condens. Matter} \textbf{\bibinfo{volume}{20}},
  \bibinfo{pages}{295211} (\bibinfo{year}{2008}).

\bibitem[{\citenamefont{Lu et~al.}(2008)\citenamefont{Lu, Beaton, Lewis,
  Tiedje, and Whitwick}}]{Lu_APL_2008}
\bibinfo{author}{\bibfnamefont{X.}~\bibnamefont{Lu}},
  \bibinfo{author}{\bibfnamefont{D.~A.} \bibnamefont{Beaton}},
  \bibinfo{author}{\bibfnamefont{R.~B.} \bibnamefont{Lewis}},
  \bibinfo{author}{\bibfnamefont{T.}~\bibnamefont{Tiedje}}, \bibnamefont{and}
  \bibinfo{author}{\bibfnamefont{M.~B.} \bibnamefont{Whitwick}},
  \bibinfo{journal}{Appl. Phys. Lett.} \textbf{\bibinfo{volume}{92}},
  \bibinfo{pages}{192110} (\bibinfo{year}{2008}).

\bibitem[{\citenamefont{Lu et~al.}(2009)\citenamefont{Lu, Beaton, Lewis,
  Tiedje, and Zhang}}]{Lu_APL_2009}
\bibinfo{author}{\bibfnamefont{X.}~\bibnamefont{Lu}},
  \bibinfo{author}{\bibfnamefont{D.~A.} \bibnamefont{Beaton}},
  \bibinfo{author}{\bibfnamefont{R.~B.} \bibnamefont{Lewis}},
  \bibinfo{author}{\bibfnamefont{T.}~\bibnamefont{Tiedje}}, \bibnamefont{and}
  \bibinfo{author}{\bibfnamefont{Y.}~\bibnamefont{Zhang}},
  \bibinfo{journal}{Appl. Phys. Lett.} \textbf{\bibinfo{volume}{95}},
  \bibinfo{pages}{041903} (\bibinfo{year}{2009}).

\bibitem[{\citenamefont{Tomi\'{c} et~al.}(2003)\citenamefont{Tomi\'{c},
  O'Reilly, Fehse, Sweeney, Adams, Andreev, Choulis, Hosea, and
  Riechert}}]{Tomic_IEEEJSTQE_2003}
\bibinfo{author}{\bibfnamefont{S.}~\bibnamefont{Tomi\'{c}}},
  \bibinfo{author}{\bibfnamefont{E.~P.} \bibnamefont{O'Reilly}},
  \bibinfo{author}{\bibfnamefont{R.}~\bibnamefont{Fehse}},
  \bibinfo{author}{\bibfnamefont{S.~J.} \bibnamefont{Sweeney}},
  \bibinfo{author}{\bibfnamefont{A.~R.} \bibnamefont{Adams}},
  \bibinfo{author}{\bibfnamefont{A.~D.} \bibnamefont{Andreev}},
  \bibinfo{author}{\bibfnamefont{S.~A.} \bibnamefont{Choulis}},
  \bibinfo{author}{\bibfnamefont{T.~J.~C.} \bibnamefont{Hosea}},
  \bibnamefont{and} \bibinfo{author}{\bibfnamefont{H.}~\bibnamefont{Riechert}},
  \bibinfo{journal}{IEEE J. Sel. Top. Quant. Electron.}
  \textbf{\bibinfo{volume}{9}}, \bibinfo{pages}{1228} (\bibinfo{year}{2003}).

\bibitem[{\citenamefont{Vurgaftman and Meyer}(2003)}]{Vurgaftman_JAP_2003}
\bibinfo{author}{\bibfnamefont{I.}~\bibnamefont{Vurgaftman}} \bibnamefont{and}
  \bibinfo{author}{\bibfnamefont{J.~R.} \bibnamefont{Meyer}},
  \bibinfo{journal}{J. Appl. Phys.} \textbf{\bibinfo{volume}{94}},
  \bibinfo{pages}{3675} (\bibinfo{year}{2003}).

\bibitem[{\citenamefont{Ferhat and Zaoui}(2006)}]{Ferhat_PRB_2006}
\bibinfo{author}{\bibfnamefont{M.}~\bibnamefont{Ferhat}} \bibnamefont{and}
  \bibinfo{author}{\bibfnamefont{A.}~\bibnamefont{Zaoui}},
  \bibinfo{journal}{Phys. Rev. B} \textbf{\bibinfo{volume}{73}},
  \bibinfo{pages}{115107} (\bibinfo{year}{2006}).

\end{thebibliography}

% These are the contents of the file 12and14band.bbl generated by BibTeX

\end{document}